\documentclass{PoS}

\usepackage[version=0.96]{pgf}
\usepackage{tikz}
\usetikzlibrary{matrix,arrows,shapes,snakes,automata,backgrounds,petri}
\usetikzlibrary{trees}
\usetikzlibrary{decorations.pathmorphing}
\usetikzlibrary{decorations.markings}

\usepackage{xspace}
\usepackage{amsmath}
\def\deltam     {\ensuremath{\Delta m}\xspace}

\def\DDbar   {\ensuremath{\kern -0.1em \stackrel{\kern 0.1em \textsf{\fontsize{5pt}{1em}\selectfont(---)}}{D}\kern -0.3em}\xspace}
\def\AAbar   {\ensuremath{\kern -0.2em \stackrel{\kern 0.2em \textsf{\fontsize{5pt}{1em}\selectfont(---)}}{A}\kern -0.3em}\xspace}
\def\AfAfbar   {\ensuremath{\kern -0.2em \stackrel{\kern 0.2em \textsf{\fontsize{5pt}{1em}\selectfont(---)}}{A}_{\kern -0.3em f}\kern -0.3em}\xspace}
\def\dacp     {\ensuremath{\Delta a_{\CP}}\xspace}
\def\CP                {\ensuremath{C\!P}\xspace}
\def\CPT               {\ensuremath{C\!P\!T}\xspace}
\def\Dbar    {\kern 0.2em\overline{\kern -0.2em \PD}{}\xspace}
\def\D       {\ensuremath{\PD}\xspace}
\def\Db      {\ensuremath{\Dbar}\xspace}
\def\Dz      {\ensuremath{\D^0}\xspace}
\def\Dzb     {\ensuremath{\Dbar^0}\xspace}
\def\DzDzb   {\ensuremath{\Dz {\kern -0.16em \Dzb}}\xspace}
\def\Dp      {\ensuremath{\D^+}\xspace}
\def\Dm      {\ensuremath{\D^-}\xspace}

\def\DpDm    {\ensuremath{\Dp {\kern -0.16em \Dm}}\xspace}
\def\Dstar   {\ensuremath{\D^*}\xspace}

\def\Dstarp  {\ensuremath{\D^{*+}}\xspace}

\def\DsJ     {\ensuremath{\D_{\squark J}}\xspace}
\def\Kstar   {\ensuremath{\PK^*}\xspace}
\def\lhcb {LHCb\xspace}

\def\babar  {BaBar\xspace}
\def\belle  {Belle\xspace}
\def\cleo   {CLEO\xspace}

\def\cdf    {CDF\xspace}
\def\PD      {\ensuremath{D}\xspace}                 
\def\PM      {\ensuremath{M}\xspace}                 
\def\PB      {\ensuremath{B}\xspace}                 
\def\Bz      {\ensuremath{\B^0}\xspace}
\def\Bzb     {\ensuremath{\Bbar^0}\xspace}
\def\Bp      {\ensuremath{\B^+}\xspace}
\def\Bm      {\ensuremath{\B^-}\xspace}
\def\Bd      {{\ensuremath{\B^0}}\xspace}
\def\Bs      {{\ensuremath{\B^0_\squark}}\xspace}
\def\PK      {\ensuremath{K}\xspace}                 
\def\kaon  {\ensuremath{\PK}\xspace}
\def\Kz    {\ensuremath{\kaon^0}\xspace}
\def\kbar    {\kern 0.2em\overline{\kern -0.2em \PK}{}\xspace}
\def\Kzb   {\ensuremath{\kbar^0}\xspace}
\def\Kp    {\ensuremath{\kaon^+}\xspace}
\def\Km    {\ensuremath{\kaon^-}\xspace}
\def\Ppi         {\ensuremath{\pi}\xspace}                 
\def\Prho        {\ensuremath{\rho}\xspace}                 
\def\pion  {\ensuremath{\Ppi}\xspace}
\def\ycp        {\ensuremath{y_{CP}}\xspace}
\def\agamma     {\ensuremath{A_{\Gamma}}\xspace}
\def\KS    {\ensuremath{\kaon^0_{\rm\scriptstyle S}}\xspace} 
\def\KL    {\ensuremath{\kaon^0_{\rm\scriptstyle L}}\xspace} 
\newcommand{\ie}{\mbox{\itshape i.e.}}
\newcommand{\eg}{\mbox{\itshape e.g.}}
\def\Bbar    {\kern 0.2em\overline{\kern -0.2em \PB}{}\xspace}
\def\B       {\ensuremath{\PB}\xspace}

\def\Jpsi     {\ensuremath{{\PJ\mskip -3mu/\mskip -2mu\Ppsi\mskip 2mu}}\xspace}
\def\PJ      {\ensuremath{\mathrm{J}}\xspace}                 
\def\Ppsi        {\ensuremath{\psi}\xspace}                 
\def\Pupsilon    {\ensuremath{\Upsilon}\xspace}                 
\def\squark    {\ensuremath{\Ps}\xspace}
\def\Ps      {\ensuremath{\mathrm{s}}\xspace}                 
\def\dquark    {\ensuremath{\Pd}\xspace}
\def\Pd      {\ensuremath{\mathrm{d}}\xspace}                 
\def\uquark    {\ensuremath{\Pu}\xspace}
\def\Pu      {\ensuremath{\mathrm{u}}\xspace}                 
\def\cquark    {\ensuremath{\Pc}\xspace}
\def\Pc      {\ensuremath{\mathrm{c}}\xspace}                 
\def\bquark    {\ensuremath{\Pb}\xspace}
\def\Pb      {\ensuremath{\mathrm{b}}\xspace}                 
\def\ubarquark    {\ensuremath{\overline{\Pu}}\xspace}
\def\cbarquark    {\ensuremath{\overline{\Pc}}\xspace}
\def\Pq      {\ensuremath{\mathrm{q}}\xspace}                 
\def\quark    {\ensuremath{\Pq}\xspace}
\def\barquark    {\ensuremath{\overline{\Pq}}\xspace}

\newcommand{\decay}[2]{\ensuremath{#1\!\to #2}\xspace}         
\def\DDbar   {\ensuremath{\kern -0.1em \stackrel{\kern 0.1em \textsf{\fontsize{5pt}{1em}\selectfont(---)}}{D}\kern -0.3em}\xspace}
\def\AAbar   {\ensuremath{\kern -0.2em \stackrel{\kern 0.2em \textsf{\fontsize{5pt}{1em}\selectfont(---)}}{A}\kern -0.3em}\xspace}
\def\AfAfbar   {\ensuremath{\kern -0.2em \stackrel{\kern 0.2em \textsf{\fontsize{5pt}{1em}\selectfont(---)}}{A}_{\kern -0.3em f}\kern -0.3em}\xspace}
\def\deltam   {\ensuremath{\delta m}\xspace}
\def\besiii {BESIII\xspace}
\def\Mbar    {\kern 0.2em\overline{\kern -0.2em \PM}{}\xspace}
\def\M       {\ensuremath{\PM}\xspace}
\def\Mz      {\ensuremath{\M^0}\xspace}
\def\Mzb     {\ensuremath{\Mbar^0}\xspace}
\def\Pe      {\ensuremath{\mathrm{e}}\xspace}                 
\def\ep         {\ensuremath{\Pe^+}\xspace}
\def\en         {\ensuremath{\Pe^-}\xspace}
\def\epem       {\ensuremath{\Pe^+\Pe^-}\xspace}
\def\mup        {\ensuremath{\Pmu^+}\xspace} 
\def\mun        {\ensuremath{\Pmu^-}\xspace} 
\def\mump       {\ensuremath{\Pmu^\mp}\xspace} 
\def\epm       {\ensuremath{\Pe^\pm}\xspace} 
\def\Pmu         {\ensuremath{\mu}\xspace}                 
\def\neu        {\ensuremath{\Pnu}\xspace}
\def\neub       {\ensuremath{\overline{\Pnu}}\xspace}

\def\neum       {\ensuremath{\neu_\mu}\xspace}
\def\neumb      {\ensuremath{\neub_\mu}\xspace}

\def\Pnu         {\ensuremath{\nu}\xspace}                 

\def\mup        {\ensuremath{\Pmu^+}\xspace}
\def\pion  {\ensuremath{\Ppi}\xspace}
\def\piz   {\ensuremath{\pion^0}\xspace}

\def\pip   {\ensuremath{\pion^+}\xspace}
\def\pipm  {\ensuremath{\pion^\pm}\xspace}

\def\pim   {\ensuremath{\pion^-}\xspace}
\def\lhcb {LHCb\xspace}
\def\ux85 {UX85\xspace}

\def\babar  {BaBar\xspace}
\def\belle  {Belle\xspace}

\def\cdf    {CDF\xspace}

\def\cleo   {CLEO\xspace}

\def\tevatron {Tevatron\xspace}
 \def\Pphi        {\ensuremath{\phi}\xspace}                 
  \def\Dbar    {\kern 0.2em\overline{\kern -0.2em \PD}{}\xspace}
\def\D       {\ensuremath{\PD}\xspace}
\def\Db      {\ensuremath{\Dbar}\xspace}
\def\Dz      {\ensuremath{\D^0}\xspace}
\def\Dzb     {\ensuremath{\Dbar^0}\xspace}
\def\DzDzb   {\ensuremath{\Dz {\kern -0.16em \Dzb}}\xspace}
\def\Dp      {\ensuremath{\D^+}\xspace}
\def\Dm      {\ensuremath{\D^-}\xspace}

\def\DpDm    {\ensuremath{\Dp {\kern -0.16em \Dm}}\xspace}
\def\Dstar   {\ensuremath{\D^*}\xspace}

\def\Dstarp  {\ensuremath{\D^{*+}}\xspace}

\def\Ds      {\ensuremath{\D^+_\squark}\xspace}
\def\Dsp     {\ensuremath{\D^+_\squark}\xspace}
\def\Dsm     {\ensuremath{\D^-_\squark}\xspace}

\def\Bd      {\ensuremath{\B^0_\dquark}\xspace}
\def\Bs      {\ensuremath{\B^0_\squark}\xspace}

\def\Wpm    {\ensuremath{\PW^\pm}\xspace}
\def\PW      {\ensuremath{W}\xspace}

\def\mbarn{\ensuremath{\rm \,mb}\xspace}
\def\mub{\ensuremath{\rm \,\mu b}\xspace}

\def\nb {\ensuremath{\rm \,nb}\xspace}

\def\invfb   {\ensuremath{\mbox{\,fb}^{-1}}\xspace}

\def\invab   {\ensuremath{\mbox{\,ab}^{-1}}\xspace}

\newcommand{\tev}{\ensuremath{\mathrm{\,Te\kern -0.1em V}}\xspace}
\newcommand{\gev}{\ensuremath{\mathrm{\,Ge\kern -0.1em V}}\xspace}
\newcommand{\mev}{\ensuremath{\mathrm{\,Me\kern -0.1em V}}\xspace}
\newcommand{\kev}{\ensuremath{\mathrm{\,ke\kern -0.1em V}}\xspace}
\newcommand{\ev}{\ensuremath{\mathrm{\,e\kern -0.1em V}}\xspace}
\newcommand{\gevc}{\ensuremath{{\mathrm{\,Ge\kern -0.1em V\!/}c}}\xspace}
\newcommand{\mevc}{\ensuremath{{\mathrm{\,Me\kern -0.1em V\!/}c}}\xspace}
\newcommand{\gevcc}{\ensuremath{{\mathrm{\,Ge\kern -0.1em V\!/}c^2}}\xspace}
\newcommand{\gevgevcccc}{\ensuremath{{\mathrm{\,Ge\kern -0.1em V^2\!/}c^4}}\xspace}
\newcommand{\mevcc}{\ensuremath{{\mathrm{\,Me\kern -0.1em V\!/}c^2}}\xspace}

\def\ps   {\ensuremath{{\rm \,ps}}\xspace}

\def\invps{\ensuremath{{\rm \,ps^{-1}}}\xspace}

\usepackage{lineno}

\title{Introduction to Charm Physics}

\ShortTitle{Charm Physics}

\author{\speaker{Marco Gersabeck}%
  \thanks{The author acknowledges support from STFC, UK, through grant ST/J004332/1.}\\
    The University of Manchester\\
    E-mail: \email{marco.gersabeck@manchester.ac.uk}
}

\abstract{
This paper gives an overview of charm physics. 
It is a lecture write-up aimed at students with a minimum of prior knowledge in particle physics, but at the same time provides a state-of-the art review of the field.
The main focus is on mixing and \CP violation, which is a field with ever growing attention since first evidence for charm mixing was observed in 2007.
Other areas covered are charm spectroscopy, production, as well as rare decays.
}

\FullConference{Flavorful Ways to New Physics\\
  28-31 October 2014\\
    Freudenstadt - Lauterbad, Germany}

\definecolor{darkyellow}{rgb}{0.7,0.7,0}
\definecolor{grey}{rgb}{0.7,0.7,0.7}

\begin{document}

\tikzset{
    vertex/.style={circle,draw=violet!50,fill=violet!20,thick},
    photon/.style={decorate, decoration={snake}, draw=darkyellow},
    electron/.style={draw=violet, postaction={decorate},
        decoration={markings,mark=at position .55 with {\arrow[draw=violet]{>}}}},
    antielectron/.style={draw=violet, postaction={decorate},
        decoration={markings,mark=at position .55 with {\arrow[draw=violet]{<}}}},
    gluon/.style={decorate, draw=magenta,
        decoration={coil,amplitude=4pt, segment length=5pt}},
    majorana/.style={draw=violet, postaction={decorate},
        decoration={markings,mark=at position .52 with {\arrow[draw=violet]{<}},mark=at position .48 with {\arrow[draw=violet]{>}}}}
}

\section{Introduction}

\noindent This paper is based on a lecture given at the workshop ``Flavorful Ways to New Physics".
However, its scope extends beyond the topics that could be covered in the limited time of the lecture.
The aim is to give a broad but at the same time sufficiently in-depth overview of the topic of charm physics to be of benefit to students with a minimum of background in particle physics.
The structure and content of the paper is based on a review article~\cite{Gersabeck:2012rp}, with several additional topics, more pedagogical detail as well as a content that has been brought up to date. 

Charm physics covers the studies of a range of composite particles containing charm quarks which provide unique opportunities for probing the strong and weak interactions in the standard model and beyond.
The charm quark, being the up-type quark of the second of the three generations, is the third-heaviest of the six quarks.
Charm particles can exist as so-called open charm mesons or baryons, containing one or several (for baryons) charm quarks, or as charmonium states which are bound states of charm anti-charm ($\cquark\cbarquark$) quark pairs.
In addition, several states containing charm quarks have been observed that cannot be explained as conventional mesons or baryons or as bound states thereof.
Their exact nature is one of today's unanswered questions.

The uniqueness of charm particles lies in their decays.
The charm quark can only decay via weak decays, mediated by a \Wpm-boson, into a strange or down quark.
An exception to this are decays of ground state charmonium mesons, which decay via annihilation of the charm and anti-charm quarks.
Thus, open charm particles are the only ones allowing the study of weak decays of an up-type quark in a bound state.

In 2009, Ikaros Bigi asked whether charm's third time could be the real charm~\cite{Bigi:2009jj}.
Charm's first time was the discovery of the \Jpsi~\cite{Aubert:1974js,Augustin:1974xw}, which followed three years after the possible first observation of an open charm decay in cosmic ray showers~\cite{Niu:1971xu}. 
This discovery confirmed the existence of a fourth quark as expected by the GIM mechanism~\cite{Glashow:1970gm} motivated by the non-existence of flavour-changing neutral currents~\cite{Bott:1967,Foeth:1969hi} in conjunction with the observation of the mixing of neutral kaons~\cite{Lande:1956pf,Jackson:1957zzb,Niebergall:1974wh}.
The second time charm attracted considerable attention was caused by the observation of \DsJ states,~\cite{Aubert:2003fg,Besson:2003cp,Abe:2003jk,Aubert:2003pe} which could not be accommodated by QCD motivated quark models~\cite{Godfrey:1985xj,Godfrey:1986wj,Isgur:1991wq,DiPierro:2001uu,Matsuki:2007zza}.
Until today, excited charmonium and open charm particles provide an excellent laboratory for studying QCD.

Charm's third time started with the first evidence for mixing of neutral charm mesons reported by \babar~\cite{Aubert:2007wf} and \belle~\cite{Staric:2007dt} in 2007.
Since then a lot of work went into more precise measurements of the mixing phenomenon as well as into searches for charge-parity (\CP) symmetry violation in the charm sector.
At the same time theoretical calculations were improved even though precise standard model predictions are still a major challenge.

In 2013, the observation of a charged charmonium-like state has marked the beginning of what might be called charm's fourth time.
The state with a mass of about $3900\mevcc$ and a width in the region of $50\mevcc$ was first observed in decays to $J/\Ppsi\pipm$ final states~\cite{Liu:2013dau,Xiao:2013iha}.
A similar state or possibly the same has been observed at the threshold of the $\PD\Dbar^*$ spectrum~\cite{Ablikim:2013emm}.
Its charmonium content and electrical charge rule out any interpretation as a two-quark or three-quark state.
However, also attempts to identify the state \eg\ as a molecule of two charm mesons or as a tetraquark state do not yield clear results.
Already in 2008, the Belle collaboration observed a similar charged charmonium-like state around a mass of $4430\mevcc$~\cite{Choi:2007wga}.
However, this could not be confirmed by a BaBar analysis~\cite{Aubert:2008aa}.
Further studies by the Belle collaboration~\cite{Mizuk:2009da,Chilikin:2013tch} gave additional insight into spin and parity of the state and supported their initial measurement.
This was finally confirmed by an LHCb analysis based on a significantly larger dataset~\cite{Aaij:2014jqa}.
Detailed reviews of these states are available in Refs.~\cite{Esposito:2014rxa,Olsen:2014qna}.

The focus of this paper lies on the current situation of studies of processes, which are mediated by the weak interaction, using open charm particles.
Particular focus is given to mixing and \CP violation, followed by comments on rare charm decays at the end of this review.

\subsection{Charm production}
\label{sec:charm_prod}

\noindent Charm physics has been and is being performed at a range of different accelerators.
These come with different production mechanisms and thus with largely varying production cross-sections.
At $\ep\en$ colliders two different running conditions are of interest to charm physics.
Tuning the centre-of-mass energy to slightly more than $3770\mevcc$ resonantly produces $\Ppsi(3770)$ mesons.
These decay almost exclusively to quantum-correlated $\Dz\Dzb$ or $\Dp\Dm$ pairs.
The use of this production type has been pioneered by the MARK III collaboration at the $\ep\en$ storage ring SPEAR at SLAC~\cite{Baltrusaitis:1985iw}.
More recently, this production mode has been used for the CLEO-c experiment at the CESR-c collider as well as for BESIII at BEPCII.
For all these experiments the collisions happen at rest, which removes the possibility to study the decay-time structure of the \PD mesons.

The most commonly used alternative is running at a higher centre-of-mass energy to produce $\Pupsilon(4S)$ on resonance which decay into quantum-correlated $\Bz\Bzb$ or $\Bp\Bm$ pairs.
This is used by the \babar and \belle experiments which are located at the PEP-II and KEKB colliders, respectively.
Both PEP-II and KEKB are asymmetric colliders, which means that the energies of the \ep and \en beams differ.
Hence, they have a collision system that is boosted with respect to the laboratory frame.
This allows measurements with decay-time resolutions about a factor two to four below the \Dz lifetime and therefore decay-time dependent studies.
Currently, the Super-KEKB collider is under construction, which will act as a \PB-factory with much improved luminosity for the Belle II experiment.

The production cross-section for producing $\D\Db$ pairs at the $\Ppsi(3770)$ resonance is approximately $8\nb$~\cite{Ablikim:2004ck}.
When running at the $\Pupsilon(4S)$ resonance, the cross-section for producing at least one \Dz meson is $1.45\nb$~\cite{Bevan:2014iga}.
The latter scenario gives access to all species of charm particles while the $\Ppsi(3770)$ only decays into $\Dz\Dzb$ or $\Dp\Dm$ pairs.
The \babar and \belle experiments have collected integrated luminosities of about $500\invfb$ and $1000\invfb$, respectively.
The Belle II experiment is expected to collect $50\invab$ between 2017 and 2022. 
CLEO-c has collected $0.5\invfb$ at the $\Ppsi(3770)$ resonance as well as around $0.3\invfb$ above the threshold for $\Dsp\Dsm$ production.
BESIII has so far collected nearly $3\invfb$ at the $\Ppsi(3770)$ resonance as well as nearly $6\invfb$ at energies between $3.9\gev$ and $4.6\gev$.

At hadron colliders the production cross-sections are significantly higher.
The cross-section for producing $\cquark\cbarquark$ pairs in proton-proton collisions at the LHC with a centre-of-mass energy $7\tev$ is about $6\mbarn$~\cite{LHCb-CONF-2010-013}, \ie\ more than six orders of magnitude higher compared to operating an $\ep\en$ collider at the $\Pupsilon(4S)$ resonance.
This corresponds to a cross-section of about $1.4\mbarn$ for producing $\Dz$ in the \lhcb acceptance\footnote{Given as a range in momentum transverse to the beam direction and rapidity as $p_{\rm T}<8\gevc,2<y<4.5$.}~\cite{Aaij:2013mga}.
This number may be compared to its equivalent at \cdf which has been measured to $13\mub$ inside the detector acceptance\footnote{The \cdf acceptance is defined as $p_{\rm T}>5.5\gevc,|y|<1$.} for proton anti-proton collisions at the \tevatron with $\sqrt{s}=1.96\tev$~\cite{Acosta:2003ax}.
\cdf has collected a total of about $10\invfb$ while \lhcb has collected $3\invfb$ during the first run of the LHC, corresponding to $0.13\times10^{12}$ and $4.6\times10^{12}$ \Dz mesons produced in the respective detector acceptances.
Table~\ref{tab:charm_prod} summarises the different charm productions with the example of the production of \Dz mesons.

\begin{table}[tb]
\centering
\begin{tabular}{lccccc}
Experiment & Year      &  $\sqrt{s}$ & $\sigma_{acc}(\Dz)$ & $L$        & $n(\Dz)$\\
\hline
CLEO-c     & 2003-2008 & $3.77\gev$  & $8\nb$           & $0.5\invfb$ & $4.0\times10^{6}$\\
BESIII     & 2010-2011 & $3.77\gev$  & $8\nb$           & $3\invfb$ & $2.4\times10^{7}$\\
\babar     & 1999-2008 & $10.6\gev$  & $1.45\nb$ & $500\invfb$ & $7.3\times10^{8}$\\
\belle     & 1999-2010 & $10.6-10.9\gev$  & $1.45\nb$ & $1000\invfb$ & $1.5\times10^{9}$\\
\cdf       & 2001-2011 & $2\tev$     & $13\mub$         & $10\invfb$ & $1.3\times10^{11}$\\
\lhcb      & 2011      & $7\tev$     & $1.4\mbarn$         & $1\invfb$  & $1.4\times10^{12}$\\
\lhcb      & 2012      & $8\tev$     & $1.6\mbarn^*$       & $2\invfb$  & $3.2\times10^{12}$
\end{tabular}
\caption{\label{tab:charm_prod}Charm production values for different experiments based on the example of the production of \Dz mesons inside the respective detector acceptances as defined in the text. The \lhcb \Dz production cross-section at $\sqrt{s}=8\tev$ has been extrapolated from that at $\sqrt{s}=7\tev$ assuming linear scaling with $\sqrt{s}$.}
\end{table}

The production of charm quarks in hadron collisions occurs predominantly in very asymmetric collisions which result in heavily boosted quarks with high rapidities.
While the asymmetry of the collisions has its origin in two different beam energies at \PB-factories, it comes from the fact that LHC collisions occur among partons rather than the protons as a whole.
At LHC energies it is mostly the gluons participating in the collisions and the fractional energy of the two colliding partons is likely to differ.
This difference leads to a smaller centre-of-mass energy as well as a boost of the collision system.
Therefore, \lhcb is ideally suited for performing decay-time dependent studies of charm decays.
At the same time, the $\cquark\cbarquark$ cross-section at the LHC is about $10\%$ of the total inelastic cross-section which allows to have reasonably low background levels for a hadronic environment.
The coverage of nearly the full solid angle of the $\ep\en$-collider experiments mentioned here makes them very powerful instruments for analysing decays involving neutral particles that may remain undetected or for inclusive studies.
In addition, the low multiplicity of collision products makes these collisions particularly clean laboratories due to the low level of backgrounds.

\subsection{Charm spectroscopy}
\label{sec:charm_spectroscopy}

\noindent Open charm mesons and baryons and charmonium mesons exhibit a rich structure of excited states.
Based on the quark model the expected spectra have been predicted thirty years ago~\cite{Godfrey:1985xj}.
The 1S ground states $\PD_{(s)}$, and $\Dstar_{(s)}$ have spin-parity $J^P=0^-$, and $1^-$, respectively, and exist for the neutral ($\cquark\bar\uquark$) and charged ($\cquark\bar\dquark$, $\cquark\bar\squark$) mesons.
In general, the configuration with $P=(-1)^J$ is called natural parity and denoted as $\Dstar$, while $P=-(-1)^J$ is called unnatural parity.
The ground state transition \decay{\Dstarp}{\Dz\pip} will be of importance for identifying the flavour of \Dz mesons later in this paper.

The first orbital excitations ($nL=1P$ multiplets) have been already observed for both charm and charm-strange mesons. 
For the latter for example, these are $\PD^*_{s0}(2317)$ ($0+$), two $1+$ states $\PD_{s1}(2460)$ and $\PD_{s1}(2536)$, as well as $\PD^*_{s2}(2573)$ with spin-parity $2+$. 
The two lightest ones were found to be much lighter and narrower than the quark models predicted.
This triggered a lively discussion about the nature of these mesons and led to the revision of the corresponding calculations. 
There are also a number of candidates for the first radial excitations ($n=2$) and/or higher orbital excitations observed.

A similarly rich spectrum exists for charmonium, \ie\ $\cquark\bar\cquark$ mesons.
The ground states are $\eta_c(1S)$ and $J/\Ppsi(1S)$ with spin-parity $J^P=0^-$, and $1^-$, respectively.
A detailed review of charmonium spectroscopy can be found in Ref.~\cite{Brambilla:2010cs}.
Thirdly, charm baryons also have a rich spectroscopy; however, this is the least explored sector as so far only singly charmed baryons have been observed.

A number of the quark model states have now been experimentally confirmed.
However, there are observed states that cannot be matched to spectroscopy predictions.
In addition to spectroscopy, the study of production and decay modes of the individual states are needed to pin down their precise nature.
More exotic models exist that attempt their explanation, such as meson molecules, quark-gluon hybrids, glueballs.
Other approaches are based particles made of more than three quarks to form tetraquarks, pentaquarks, or H-dibaryons.

\subsection{Mixing of neutral mesons}
\label{sec:mix_intro}

\noindent For neutral mesons, the mass eigenstates, \ie\ the physical particles of defined mass and lifetime, do not {\it a priori} coincide with the flavour eigenstates, which are being produced and evolve in time into their anti-particles to finally decay as mass eigenstates.
In general the state of a meson can be expressed as a linear combination of the flavour eigenstates as
\begin{equation}
|\psi\rangle=a(t)|\Mz\rangle + b(t)|\Mzb\rangle,
\end{equation}
which is conveniently abbreviated to
\begin{equation}
|\psi\rangle=\left(\begin{array}{c}a(t)\\b(t)\end{array}\right).
\end{equation}
For the purpose of mixing it is useful to apply the Weisskopf-Wigner approximation~\cite{Weisskopf:1930ps,Weisskopf:1930au} by neglecting the flavour-conserving strong and electromagnetic interactions and writing the Schr\"odinger equation
\begin{equation}
\imath\hbar\frac{\partial}{\partial t}\psi={\mathcal H}\psi,
\end{equation}
with the effective Hamiltonian given by
\begin{equation}
{\mathcal H}=M-\frac{\imath}{2}\Gamma=\left(\begin{array}{cc}
M_{11}-\frac{\imath}{2}\Gamma_{11} & M_{12}-\frac{\imath}{2}\Gamma_{12}\\
M_{21}-\frac{\imath}{2}\Gamma_{21} & M_{22}-\frac{\imath}{2}\Gamma_{22}\end{array}\right),
\end{equation}
where the mass matrix $M$ and the decay matrix $\Gamma$ are Hermitian.
The eigenstates of the Hamiltonian, $|M_{1,2}\rangle$, have eigenvalues
\begin{equation}
\lambda_{1,2}\equiv m_{1,2}-\frac{\imath}{2}\Gamma_{1,2},
\end{equation}
as well as the observable masses $m_{1,2}$ and decay widths $\Gamma_{1,2}$. 
The time evolution of the physical states is therefore given as
\begin{equation}
|M_{1,2}(t)\rangle=e^{-\imath m_{1,2}t}e^{-\Gamma_{1,2}t/2}|M_{1,2}(0)\rangle.
\label{eqn:m12t}
\end{equation}
Assuming \CPT symmetry, the physical eigenstates can be expressed as
\begin{equation}
|\PM_{1,2}\rangle=p|\Mz\rangle\pm{}q|\Mzb\rangle,
\label{eqn:m12}
\end{equation}
with complex coefficients satisfying $|p|^2+|q|^2=1$.
The choice of the phase convention of $p$ and $q$ can be made such that, in the limit of no \CP violation, \ie\ $|q/p|=1$, $\CP|\Mz\rangle=-|\Mzb\rangle$.
This is equivalent to stating that $|M_1\rangle$ is \CP odd and $|M_2\rangle$ is \CP even.

Solving Eq.~\ref{eqn:m12} for the flavour eigenstates one can express their time dependence by inserting Eq.~\ref{eqn:m12t} and using Eq.~\ref{eqn:m12} for the initial states to obtain
\begin{eqnarray}
|\Mz(t)\rangle & = & f_+(t)|\Mz\rangle+\frac{q}{p}f_-(t)|\Mzb\rangle\nonumber\\
|\Mzb(t)\rangle & = & f_+(t)|\Mzb\rangle+\frac{p}{q}f_-(t)|\Mz\rangle,
\label{eqn:m0t}
\end{eqnarray}
with
\begin{equation}
f_\pm(t)=\frac{1}{2}e^{-\imath m_1}e^{-\Gamma_1 t/2}\left(1\pm e^{-\imath\Delta m t}e^{\Delta\Gamma t/2}\right),
\end{equation}
where the quantities $\Delta{}m\equiv{}m_2-m_1$ and $\Delta\Gamma\equiv\Gamma_2-\Gamma_1$ have been introduced.
It is useful to introduce further the corresponding averages $m\equiv(m_1+m_2)/2$ and $\Gamma\equiv(\Gamma_1+\Gamma_2)/2$ as well as the dimensionless quantities $x\equiv\Delta{}m/\Gamma$ and $y\equiv\Delta\Gamma/(2\Gamma)$.

At this stage the transition probability of one flavour eigenstate to a state of the same or opposite flavour can be calculated as
\begin{equation}
\begin{array}{ll}
P(\decay{\Mz(t)}{\Mz}) = P(\decay{\Mzb(t)}{\Mzb}) & = |f_+(t)|^2 =  \frac{1}{2}e^{-\Gamma{}t}(\cosh(y\Gamma{}t)+\cos(x\Gamma{}t)),\\
P(\decay{\Mz(t)}{\Mzb}) & = \left|\frac{q}{p}\right|^2|f_-(t)|^2 = \frac{1}{2}\left|\frac{q}{p}\right|^{2}e^{-\Gamma{}t}(\cosh(y\Gamma{}t)-\cos(x\Gamma{}t)),\\
P(\decay{\Mzb(t)}{\Mz}) & = \left|\frac{p}{q}\right|^2|f_-(t)|^2 = \frac{1}{2}\left|\frac{p}{q}\right|^{2}e^{-\Gamma{}t}(\cosh(y\Gamma{}t)-\cos(x\Gamma{}t)).
\end{array}
\label{eqn:mixing}
\end{equation}
These equations show that it is the mixing parameters $x$, and $y$, which define the characteristic behaviour of neutral meson mixing.
In particular it is the mass difference of the physical eigenstates, which determines the rate of oscillations between mesons and anti-mesons.
Different decay widths lead to a non-oscillating change of the exponential decay.
A more detailed introduction can be found \eg\ in Refs.~\cite{Sozzi:2008zza,Bigi:2009}.

\begin{table}[tb]
\centering
\begin{tabular}{lccccc}
Meson & Mass in \gev & Width in \invps & Lifetime in \ps & x & y\\
\hline
\Kz   & $0.49761(2)$ & $0.005594(2)$ & $\begin{array}{c}89.54(4)\\51160(210)\end{array}$ & $0.946(2)$ & $0.997(1)$\\
\Dz   & $1.86484(7)$ & $2.438(9)$ & $0.4101(15)$ & $0.0039(^{+17}_{-18})$ & $0.0065(^{+07}_{-09})$\\
\Bd   & $5279.58(17)$ & $0.658(2)$ & $1.519(5)$ & $0.774(6)$ & $<0.09$\\
\Bs   & $5366.77(24)$ & $0.869(4)$ & $1.512(7)$ & $26.85(13)$ & $0.069(6)$\\
\end{tabular}
\caption{\label{tab:mixing_pars}Overview of parameters relevant to meson mixing. Masses and widths are the average of the two physical eigenstates. The uncertainties are given in brackets as multiples of the least significant digit. The values used are taken from Ref.~\cite{Agashe:2014kda}.}
\end{table}

\begin{figure}
\centering
\includegraphics[width=1.0\textwidth]{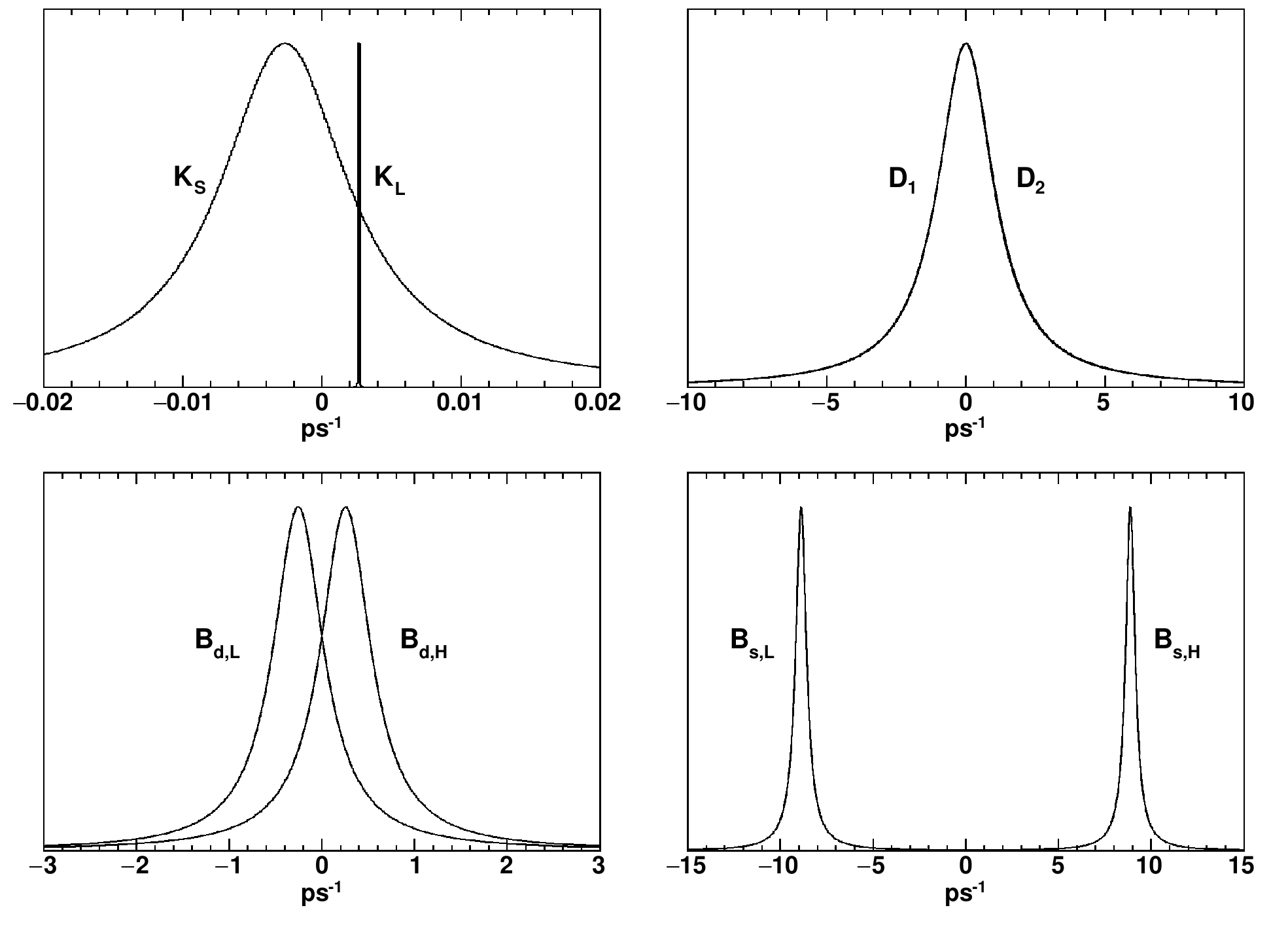}
\caption{The widths and mass differences of the physical states of the flavoured neutral mesons. The width corresponds to the inverse lifetime while the mass difference determines the oscillation frequency. The values used are taken from Ref.~\cite{Agashe:2014kda}.}
\label{fig:mesons}
\end{figure}

The four systems, which are subject to mixing, are kaons (\PK), charm (\PD), \Bd, and \Bs mesons.
For charm mesons the mixing parameters are drastically different compared to those of kaons or \PB mesons.
Their masses, lifetimes, and mixing parameters are shown in Table~\ref{tab:mixing_pars}.
Figure~\ref{fig:mesons} gives a graphical representation of the widths and mass differences of the four neutral meson systems.
The kaon system is the only one to have $y\approx 1$, resulting in two mass eigenstates with vastly different lifetimes, hence their names \PK-short (\KS) and \PK-long (\KL).
Furthermore, also $x\approx1$ which results in a sizeable sinusoidal oscillation frequency as shown in Eq.~\ref{eqn:mixing}).
The two \PB-meson systems have reasonably small width splitting; however, they have sizeable values for $x$.
Particularly for the \Bs system this leads to fast oscillations which require high experimental accuracy to be resolved.
Similarly, only recent measurements were able to measure the \Bs width splitting to be unambiguously non-zero.
The charm meson system is the only one where both $x$ and $y$ are significantly less than $1$, hence the nearly overlapping curves in Fig.~\ref{fig:mesons}.

Experimentally, the different mixing parameters lead to rather different challenges for measurements in the various meson systems.
The vast lifetime difference in the kaon system leads to the possibility of studying nearly clean samples of just one of the two mass eigenstates by either measuring decays close to a production target where \KS decays dominate, or far away where most \KS have decayed before entering the detection region.
In the \PB systems the oscillation frequency puts a challenge to the decay-time resolution, particularly for \Bs mesons as mentioned before.
The smallness of $y$ requires, to first order, large data samples to acquire the necessary statistical precision for measuring such a small quantity.
The latter is particularly true for the charm system, where both $x$ and $y$ are small.
This is the reason why it was only in 2007 when first evidence for charm mixing was observed and took until 2013 for mixing to be observed with high statistical significance in a single measurement.

\subsection{\CP violation}
\label{sec:cpv_intro}

\noindent The symmetry under \CP transformation, \ie\ the exchange of particles and anti-particles can be violated in different ways.
One is a difference in the transition probability of mesons to anti-mesons compared to the reverse process, \ie\ $P(\decay{\Mz(t)}{\Mzb})\neq P(\decay{\Mzb(t)}{\Mz})$.
As can be seen in Eq.~\ref{eqn:mixing}, this is the case if $|q/p|\neq 1$.
This type of \CP violation is independent of the decay mode.

When considering also the decay of a meson, \CP violation is given by the inequality of the magnitude of the decay amplitude of a meson to a final state $f$, $|A_f|$, compared to the magnitude of the anti-meson decay amplitude to the \CP conjugate final state $\bar{f}$, $|\bar{A}_{\bar{f}}|$.
Hence, \CP violation is present when $|\bar{A}_{\bar{f}}/A_f|\neq1$.
This decay-mode dependent \CP violation is called direct \CP violation, with the corresponding parameter defined as
\begin{equation}
a_{\rm d}\equiv\frac{|A_f|^2-|\bar{A}_{\bar{f}}|^2}{|A_f|^2+|\bar{A}_{\bar{f}}|^2}.
\end{equation}

Finally, for neutral meson decays the mixing and decay amplitudes can interfere, which can lead to a third type of \CP violation.
This \CP violation in the interference is present if the complex quantity $\lambda_f$, defined as
\begin{equation}
\label{eqn:charm_lambda}
\lambda_f\equiv\frac{q\bar{A}_{\bar{f}}}{pA_f}=-\eta_{\CP}\left|\frac{q}{p}\right|\left|\frac{\bar{A}_f}{A_f}\right|e^{i\phi},
\end{equation}
has a non-zero imaginary part.
The right-hand expression is valid for a \CP eigenstate $f$ with eigenvalue $\eta_{\CP}$ and $\phi$ is the \CP violating relative phase between $q/p$ and $\bar{A}_f/A_f$.
The combination of \CP violation in interference and in mixing is also called indirect \CP violation.
In general, it can be stated that \CP symmetry is violated if $\lambda_f\neq 1$.
An excellent discussion on the different types of \CP violation can be found in section 7.2.1 of Ref.~\cite{Sozzi:2008zza}.
As opposed to the strange and the beauty system, \CP violation has not yet been discovered in the charm system.

The occurrence of \CP violation generally requires the interference of several amplitudes contributing to a particular process, which follows from writing a decay amplitude as
\begin{equation}
A_f=C\left(1+r_fe^{\imath(\delta_f+\phi_f)}\right),
\end{equation}
where the amplitude has been split in a leading and sub-leading part with a relative magnitude $r_f$ and relative strong and weak phases $\delta_f$ and $\phi_f$, respectively.
Correspondingly, the \CP-conjugate decay is given by
\begin{equation}
\bar{A}_{\bar{f}}=C\left(1+r_fe^{\imath(\delta_f-\phi_f)}\right),
\end{equation}
where the weak phase changes sign and strong phase does not.
Using the time-integrated decay rate
\begin{equation}
\Gamma(\decay{\PM}{f})=\int_0^\infty\Gamma(\decay{\PM(t)}{f})dt\propto\left|A_f\right|^2,
\end{equation}
one obtains for the \CP asymmetry
\begin{equation}
a_{\CP}\equiv\frac{\Gamma(\decay{\PM}{f})-\Gamma(\decay{\Mbar}{\bar{f}})}{\Gamma(\decay{\PM}{f})+\Gamma(\decay{\Mbar}{\bar{f}})}=2r_f\sin\delta_f\sin\phi_f.
\end{equation}
This equation shows that \CP violation can only occur in the case that two (or more) amplitudes contribute to the decay ($r_f\neq0$) and that they have different strong and weak phases.

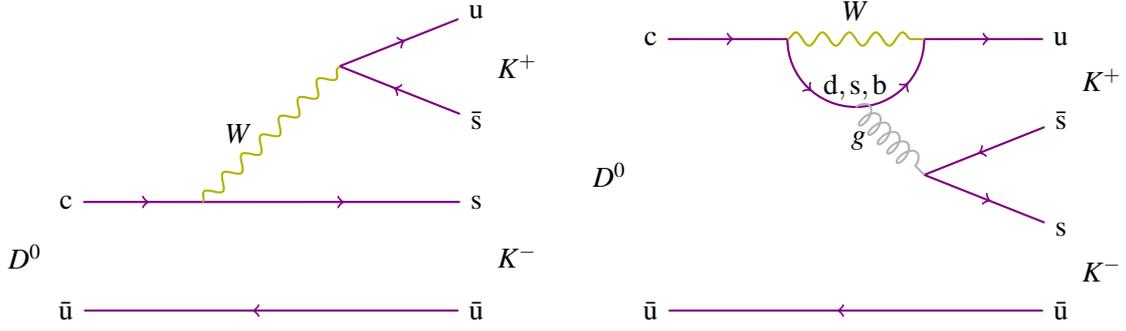
\begin{figure}
\centering
\begin{tikzpicture}[scale=1.8,thick]
\node (charm) at (0,1) [] {\cquark};
\node (strange) at (3,1) [] {\squark};
\node (antiup) at (0,0.2) [] {$\bar{\uquark}$};
\node (antiup2) at (3,0.2) [] {$\bar{\uquark}$};
\node (antistrange) at (3,1.6) [] {$\bar{\squark}$};
\node (up) at (3,2.4) [] {\uquark};
\node (Dzero) at (-0.3,0.6) [] {\Dz};
\node (Kminus) at (3.3,0.6) [] {\Km};
\node (Kplus) at (3.3,2.0) [] {\Kp};
\path[-,decoration={markings,mark=at position .55 with {\arrow[draw=violet]{>}}}]
  (charm) edge[draw=violet,postaction={decorate}] (1,1)
  (1,1) edge[draw=violet,postaction={decorate}] (strange)
  (antiup2) edge[draw=violet,postaction={decorate}] (antiup)
  (antistrange) edge[draw=violet,postaction={decorate}] (2,2.0)
  (2,2.0) edge[draw=violet,postaction={decorate}] (up);
\path[-,decoration={snake}]
  (1,1) edge[decorate,draw=darkyellow] node [left=3pt] {\PW} (2,2.0);
\end{tikzpicture}
\quad
\begin{tikzpicture}[scale=1.8,thick]
\node (charm) at (0,2.4) [] {\cquark};
\node (up) at (3,2.4) [] {\uquark};
\node (antiup) at (0,0.4) [] {$\bar{\uquark}$};
\node (antiup2) at (3,0.4) [] {$\bar{\uquark}$};
\node (antistrange) at (3,1.8) [] {$\bar{\squark}$};
\node (strange) at (3,1.0) [] {\squark};
\node (Dzero) at (-0.3,1.4) [] {\Dz};
\node (Kminus) at (3.3,0.70) [] {\Km};
\node (Kplus) at (3.3,2.10) [] {\Kp};
\path[-,decoration={markings,mark=at position .55 with {\arrow[draw=violet]{>}}}]
  (charm) edge[draw=violet,postaction={decorate}] (1,2.4)
  (1,2.4) edge [draw=violet,bend right=45,postaction={decorate}] node [right=2pt] {$\dquark,\squark,\bquark$} (1.5,1.9)
  (1.5,1.9) edge [draw=violet,bend right=45,postaction={decorate}] (2.0,2.4)
  (2,2.4) edge[draw=violet,postaction={decorate}] (up)
  (antiup2) edge[draw=violet,postaction={decorate}] (antiup)
  (2,1.4) edge[draw=violet,postaction={decorate}] (strange)
  (antistrange) edge[draw=violet,postaction={decorate}] (2,1.4);
\path[-,decoration={snake}]
  (1,2.4) edge[decorate,draw=darkyellow] node [above=3pt] {\PW} (2,2.4);
\path[-,decoration={coil,amplitude=4pt, segment length=5pt}]
  (1.5,1.9) edge[decorate,draw=grey] node [left=5pt] {$g$} (2,1.4);
\end{tikzpicture}
\caption{(Left) Tree-level diagram and (right) penguin diagram for a \decay{\Dz}{\Km\Kp} decay.}
\label{fig:intro_tree_penguin}
\end{figure}

The two amplitudes required for \CP violation can for example be given by a tree-level decay and a penguin contribution (see Fig.~\ref{fig:intro_tree_penguin}).
Penguin diagrams contain a $\quark\barquark$ pair of the same flavour, which reduces the number of decays to which they can contribute.
For charm decays, these are the singly Cabibbo-suppressed $\decay{\cquark}{\uquark\bar{\dquark}\dquark}$ and $\decay{\cquark}{\uquark\bar{\squark}\squark}$ transitions, which occur for example in the decays \decay{\Dz}{\pim\pip} and \decay{\Dz}{\Km\Kp} (see Sec.~\ref{sec:cpv_direct_meas}).

Tree-level and penguin amplitudes generally have different strong phases and the different CKM elements present in the process lead to different weak phases, thus enabling \CP violation.
An alternative to two amplitudes interfering for the same decay exists for neutral mesons.
In this case a final state can potentially be reached either through a direct decay or through the transition of the meson into its anti-meson followed by the anti-meson decay to the same final state.
Here, the difference in the weak phase can originate in the mixing process, while the meson and anti-meson decay amplitudes may have different strong phases.
An example of this scenario is the doubly Cabibbo-suppressed decay \decay{\Dz}{\Kp\pim} (see Sec.~\ref{sec:mixing_measurements} and Sec.~\ref{sec:cpv_wskpi}).

\section{Charm mixing}
\label{sec:mix_charm}
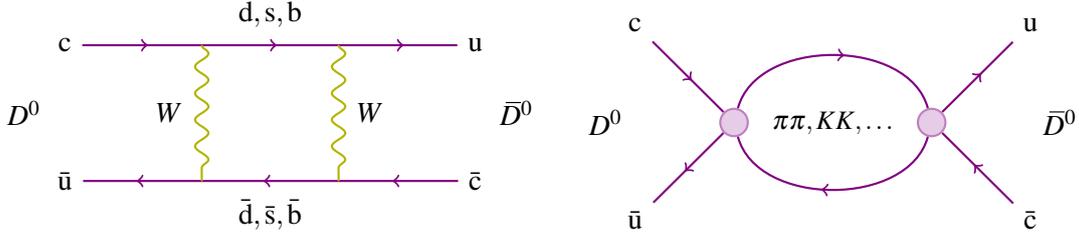
\begin{figure}
\centering
\begin{tikzpicture}[scale=1.8,thick]
\node (charm) at (0,1) [] {\cquark};
\node (up) at (3,1) [] {\uquark};
\node (antiup) at (0,0) [] {$\bar{\uquark}$};
\node (anticharm) at (3,0) [] {$\bar{\cquark}$};
\node (Dzero) at (-0.3,0.5) [] {\Dz};
\node (Dzerobar) at (3.3,0.5) [] {\Dzb};
\path[-,decoration={markings,mark=at position .55 with {\arrow[draw=violet]{>}}}]
  (charm) edge[draw=violet,postaction={decorate}] (1,1)
  (1,1) edge[draw=violet,postaction={decorate}] node [above=3pt] {$\dquark,\squark,\bquark$} (2,1)
  (2,1) edge[draw=violet,postaction={decorate}] (up)
  (anticharm) edge[draw=violet,postaction={decorate}] (2,0)
  (2,0) edge[draw=violet,postaction={decorate}] node [below=3pt] {$\bar{\dquark},\bar{\squark},\bar{\bquark}$} (1,0)
  (1,0) edge[draw=violet,postaction={decorate}] (antiup);
\path[-,decoration={snake}]
  (2,1) edge[decorate,draw=darkyellow] node [right=3pt] {\PW} (2,0)
  (1,1) edge[decorate,draw=darkyellow] node [left=3pt] {\PW} (1,0);
\end{tikzpicture}
\quad
\begin{tikzpicture}[scale=1.3,thick]
\node (charm) at (0,2) [] {\cquark};
\node (antiup) at (0,0) [] {$\bar{\uquark}$};
\node (decay) at (1,1) [vertex] {};
\node (invdecay) at (3,1) [vertex] {};
\node (up) at (4,2) [] {\uquark};
\node (anticharm) at (4,0) [] {$\bar{\cquark}$};
\node (intermediate) at (2,1) [] {\small $\pi\pi,\PK\PK,\ldots$};
\node (Dzero) at (-0.3,1) [] {\Dz};
\node (Dzerobar) at (4.3,1) [] {\Dzb};
\path[-,decoration={markings,mark=at position .55 with {\arrow[draw=violet]{>}}}]
  (charm) edge[draw=violet,postaction={decorate}] (decay)
  (decay) edge [draw=violet,bend left=75,postaction={decorate}] (invdecay)
  (invdecay) edge[draw=violet,postaction={decorate}] (up)
  (anticharm) edge[draw=violet,postaction={decorate}] (invdecay)
  (invdecay) edge [draw=violet,bend left=75,postaction={decorate}] (decay)
  (decay) edge[draw=violet,postaction={decorate}] (antiup);
\end{tikzpicture}
\caption{(Left) Box diagram and (right) re-scattering diagram for \Dz-\Dzb mixing.}
\label{fig:mix_box}
\end{figure}

\noindent The studies of charm mesons have gained in momentum with the measurements of first evidence for meson anti-meson mixing in neutral charm mesons in 2007~\cite{Aubert:2007wf,Staric:2007dt}.
Mixing is a process that changes the flavour quantum number by two units, \eg\ charmness from $+1$ to $-1$.
The only flavour-changing interaction is the weak interaction, where the $W^\pm$ boson permits $\Delta F=1$ transitions.
Thus, the only diagram for a $\Delta F=2$ process is a box diagram where two $W^\pm$ bosons are exchanged (see Fig.~\ref{fig:mix_box}).
This process is often referred to as short-distance effects in literature due to the small distance scales necessary for $W^\pm$ boson exchange.

Mixing of \Dz mesons is the only mixing process where down-type quarks contribute to the box diagram.
Unlike \PB-meson mixing, where the top-quark contribution dominates, the third generation quark is of similar mass to the other down-type quarks.
This leads to a combination of GIM cancellation~\cite{Glashow:1970gm} and CKM suppression~\cite{Cabibbo:1963yz,Kobayashi:1973fv}, which results in a strongly suppressed mixing process~\cite{Georgi:1992as,Ohl:1992sr,Bigi:2000wn,Chen:2007zua,Bobrowski:2010xg}.

Another way for neutral meson mixing to proceed is through re-scattering effects.
This can be seen as the decay of a neutral meson into a final state that is common to both flavours, followed by a recombination of the final-state particles to the anti-meson (see Fig.~\ref{fig:mix_box}).
These so-called long distance effects are likely to dominate charm meson mixing as they do not suffer from the same cancellations as the short distance effects.

There are two approaches for theoretical calculations of charm mixing.
The ``inclusive'' approach is based an operator product expansion (OPE) in the ratio of the typical hadronic scale over the charm quark mass, $\Lambda/m_{\cquark}$~\cite{Bobrowski:2010xg,Khoze:1983yp,Shifman:1984wx,Chay:1990da,Bigi:1992su,Blok:1993va,Manohar:1993qn,Lenz:2013aua}.
Due to the cancellations mentioned above it is higher order operators that give the largest contributions to the mixing parameters.
Furthermore, it is not yet clear whether the expansion series really converges since $\Lambda/m_{\cquark}\approx0.25$.
Calculations of the charm meson lifetimes are being performed to test whether the OPE approach can properly reproduce the large difference between the \Dz and the \Dp lifetimes~\cite{Lenz:2013aua}.
In the \Bs system, the OPE approach successfully predicted the width splitting of the two \Bs mass eigenstates~\cite{Lenz:2011ti} which has recently been confirmed by an \lhcb measurement~\cite{LHCb-PAPER-2013-002}.

At this point it is useful to insert a comment on \PD lifetimes.
The large ratio of the \Dp to the \Dz lifetime of about $2.5$ indicates that the spectator quark plays an important role in the calculation of hadronic decay rates of charm mesons.
In the OPE, terms depending on the spectator quark enter only at order $(\Lambda/m_{\cquark})^3$ underlining the importance of higher order calculations.
The spectator quark is the only difference in the Cabibbo-favoured \decay{\cquark}{\squark\bar{\dquark}\uquark} transitions.
For \Dp decays the spectator quark, $\bar{\dquark}$, is the same as one occurring in the charm quark decay.
Thus, there are two possible tree-level amplitudes, a colour-allowed and a colour-suppressed one, that can interfere.
Their destructive interference leads to a large suppression of the partial widths of hadronic \Dp decays compared to their \Dz counterparts.

Similarly to the decay width of a particle being given by the sum of all partial decay widths, the width splitting of the mass eigenstates can be calculated by considering the appropriate decay processes. 
The ``exclusive'' approach sums over intermediate hadronic states, taking input from models or experimental data~\cite{Donoghue:1985hh,Wolfenstein:1985ft,Colangelo:1990hj,Kaeding:1995zx,Anselm:1979fa,Falk:1999ts,Cheng:2010rv,Gronau:2013mza}.
Also in this approach, different modes of the same $SU(3)$ multiplet, \eg\ \decay{\Dz}{hh'} with $h=\PK,\pi$, lead to cancellations which is why their individual contributions have to be known to high precision.
Due to the considerable mass of the \Dz meson, many different modes need to be taken into account simultaneously.
Of these, only phase space differences can be evaluated at the moment.
Estimates indicate that mixing in the experimentally observed range is conceivable when taking into account $SU(3)$-breaking effects.
However, neither the inclusive nor the exclusive approach have thus far permitted a precise theoretical calculation of charm mixing.
A recent review of these calculations of charm mixing is given in Ref.~\cite{Petrov:2013usa}.

It was discussed whether the measured size of the mixing parameters could be interpreted as a hint for physics beyond the standard model~\cite{Hou:2006mx,Ciuchini:2007cw,Nir:2007ac,Blanke:2007ee,He:2007iu,Chen:2007yn,Golowich:2007ka}.
The biggest problem in answering this question is the non-existence of a precise standard model calculation.
Effects of physics beyond the standard model were also searched for in numerous \CP violation measurements and searches for rare decays both of which are covered in the remainder of this review.

Given the smallness of the mixing parameters $x$ and $y$ in the charm system, it is instructive to expand the terms in Eq.~\ref{eqn:mixing} to obtain
\begin{equation}
\begin{array}{ll}
P(\decay{\Mz(t)}{\Mz}) = P(\decay{\Mzb(t)}{\Mzb}) & \approx \frac{1}{2}e^{-\Gamma{}t}\left(2-\frac{(x^2-y^2)}{2}(\Gamma t)^2\right),\\
P(\decay{\Mz(t)}{\Mzb}) & \approx \frac{1}{2}\left|\frac{q}{p}\right|^{2}e^{-\Gamma{}t}\frac{(x^2+y^2)}{2}(\Gamma t)^2,\\
P(\decay{\Mzb(t)}{\Mz}) & \approx \frac{1}{2}\left|\frac{p}{q}\right|^{2}e^{-\Gamma{}t}\frac{(x^2+y^2)}{2}(\Gamma t)^2,
\end{array}
\label{eqn:mixing_charm}
\end{equation}
where terms of order $(x\Gamma t)^4$ and $(y\Gamma t)^4$ have been neglected.
Assuming a mixing parameter of the size of $0.5\%$, inspired by the latest world average~\cite{Amhis:2014hma}, and a decay time corresponding to $20$ \Dz lifetimes, this approximation corresponds to a relative error of $1\%$.

\subsection{Experimental measurements}
\subsubsection{Flavour tagging}
\noindent Mixing of \Dz mesons can be measured in several different modes.
Most require identifying the flavour of the \Dz at production as well as at the time of the decay.
Tagging the flavour at production usually exploits the strong decay \decay{\Dstarp}{\Dz\pip} (and charge conjugate\footnote{Charge conjugate decays are implicitly included henceforth.}) where the charge of the pion determines the flavour of the \Dz.
The small amount of free energy in this decay leads to the difference in the reconstructed invariant mass of the \Dstarp and the \Dz, $\deltam\equiv{}m_{\Dz\pip}-m_{\Dz}$, exhibiting a sharply peaking structure over a threshold function as background.
While the small amount of energy transfer leads to a sharply peaking signal and thus to a high signal purity, the pion from the \Dstarp decay has little kinetic energy and is thus commonly referred to as the soft pion.
This is particularly important as low momentum particles can be swept outside detector acceptances by magnetic fields, thus reducing the overall detection efficiency.
In addition, soft pions are more likely to be affected by detection asymmetries, which in the case of a tagging particle, can translate into a fake physics asymmetry.

An alternative to using the \Dstarp decay mode is tagging the \Dz flavour by reconstructing a flavour-specific decay of a \PB meson.
At the \lhcb experiment this approach is of interest due to differences in trigger efficiencies compensating for lower production rates.
In particular the decay chain \decay{\Bbar}{\Dz\mun X}, with $X$ representing a non-reconstructed fraction of the final state, is very useful due to a high branching fraction of semi-leptonic decays and due to a very high trigger efficiency for muons compared to hadrons.
Another advantage of this tagging approach is that the efficiency for reconstructing the \Dz is independent of its flight distance with respect to its origin, while \Dz mesons originating directly in the $pp$ interaction tend to have very low efficiency at low decay times.
Thus, the two tagging methods not only provide complementary datasets but also improve the decay-time coverage, which improves the lever arm for measuring time-dependent effects.

A third option available particularly at $\epem$ colliders is the reconstruction of the opposite side charm meson in a flavour specific decay.
This is based on the fact that quarks have to be produced in quark anti-quark pairs and thus every charm meson has to be accompanied by a hadron containing an anti-charm quark.
For $\epem$ colliders operating at the threshold for producing $\psi(3770)$ states the correlation goes further and the final state are quantum-entangled \Dz-\Dzb or \Dp-\Dm pairs.

\subsubsection{Mixing measurements}
\label{sec:mixing_measurements}
{\noindent \bf Measuring the mixing rate: } Theoretically, the most straight-forward mixing measurement is that of the rate of the forbidden decay \decay{\Dz}{\Kp\mun\neumb} which is only accessible through \Dz-\Dzb mixing.
The ratio of the time-integrated rate of these forbidden decays to their allowed counterparts, \decay{\Dz}{\Km\mup\neum}, thus represents the rate at which \Dz mixing occurs.
It is given, following Eq.~\ref{eqn:mixing_charm}, by
\begin{equation}
\frac{\int_0^\infty \Gamma(\decay{\Dz(t)}{\Kp\mun\neumb}) dt}{\int_0^\infty \Gamma(\decay{\Dz(t)}{\Km\mup\neum}) dt} = \frac{\int_0^\infty P(\decay{\Dz(t)}{\Dzb})|\bar{A}_{\bar{f}}|^2 dt}{\int_0^\infty P(\decay{\Dz(t)}{\Dz})|A_f|^2 dt} \approx \left|\frac{q}{p}\right|^{2}\frac{x^2+y^2}{2} \equiv \left|\frac{q}{p}\right|^{2}R_m,
\label{eqn:sl_mix}
\end{equation}
where $|A_f|^2=|\bar{A}_{\bar{f}}|^2$ are the decay rates of the decay \decay{\Dz}{\Km\mup\neum} and its \CP-conjugate, respectively, which cancel in the ratio.
Furthermore, the mixing rate $R_{\rm m}\equiv(x^2+y^2)/2$ has been introduced.

As this requires very large samples of \Dz mesons no measurement has thus far reached sufficient sensitivity to see evidence for \Dz mixing.
The most sensitive measurement to date has been made by the \belle collaboration~\cite{Bitenc:2008bk} to $R_{\rm m}=(1.3\pm2.2\pm2.0)\times10^{-4}$, where the first uncertainty is of statistical and the second is of systematic nature\footnote{This notation is applied to all results where two uncertainties are quoted.}.
Using the current world average values~\cite{Amhis:2014hma}, this can be compared to the expected rate of approximately $R_{\rm m}=(0.3\pm0.1)\times10^{-4}$.

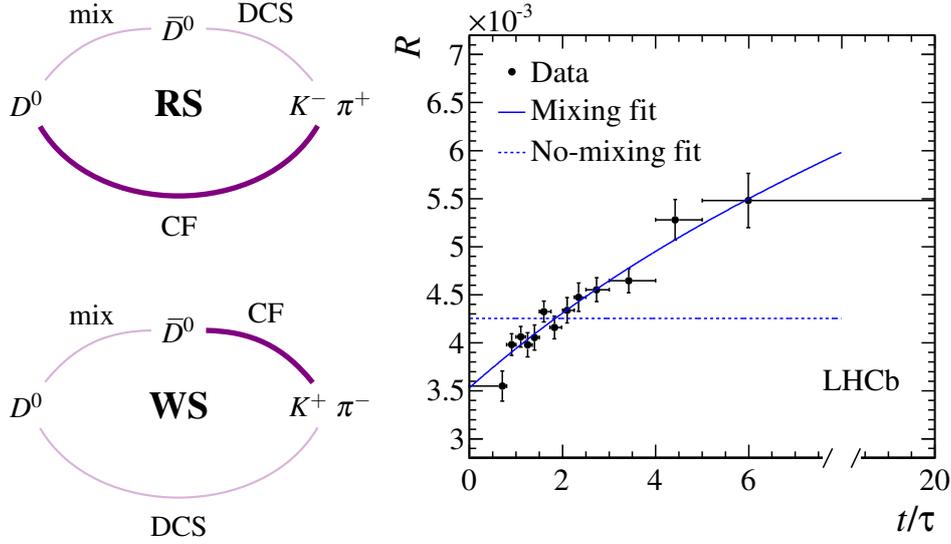
\begin{figure}
\centering
\begin{tikzpicture}[scale=2.0,thick]
\node (Dzero2) at (0,3) [] {\Dz};
\node (Dzerobar2) at (1,3.5) [] {\Dzb};
\node (Kmpip) at (2,3) [] {\Km\pip};
\node (RS) at (1,3) [] {\Large \bf RS};
\path[-]
  (Dzero2) edge [draw=violet,bend right=60, line width=2pt] node [below=3pt] {CF} (Kmpip)
  (Dzero2) edge [draw=violet!30,bend left=25] node [above=3pt] {mix} (Dzerobar2)
  (Dzerobar2) edge [draw=violet!30,bend left=25] node [above=3pt] {DCS} (Kmpip);

\node (Dzero) at (0,1) [] {\Dz};
\node (Dzerobar) at (1,1.5) [] {\Dzb};
\node (Kppim) at (2,1) [] {\Kp\pim};
\node (WS) at (1,1) [] {\Large \bf WS};
\path[-]
  (Dzero) edge [draw=violet!30,bend right=60] node [below=3pt] {DCS} (Kppim)
  (Dzero) edge [draw=violet!30,bend left=25] node [above=3pt] {mix} (Dzerobar)
  (Dzerobar) edge [draw=violet,bend left=25, line width=2pt] node [above=3pt] {CF} (Kppim);
\end{tikzpicture}
\includegraphics[width=0.5\textwidth]{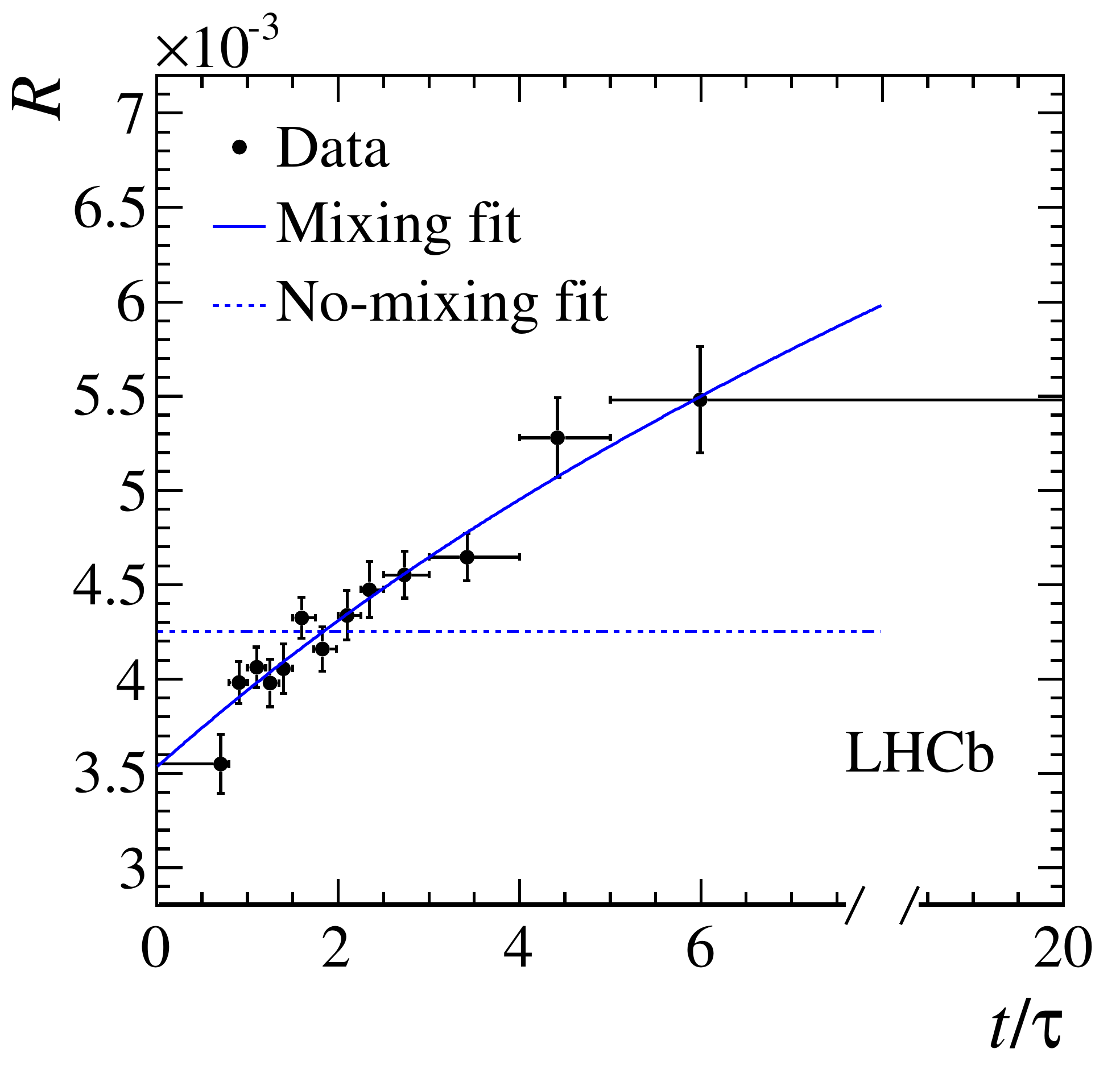}
\caption{(Left) Schematic of the amplitudes contributing to RS and WS \decay{\Dz}{\PK^\pm\pi^\mp} decays. The bold lines indicate favoured amplitudes, while the thin connections represent suppressed amplitudes. (Right) Decay-time distribution of the ratio of WS to RS \decay{\Dz}{\PK^\pm\pi^\mp} decays leading to the first single-measurement observation of charm mixing. Reproduced from Ref.~\cite{LHCb-PAPER-2012-038}.}
\label{fig:mix_dicovery}
\end{figure}

{~\\ \noindent \bf Mixing through interference: }
Related to the semi-leptonic decay is the decay \decay{\Dz}{\Kp\pim}.
This decay can proceed either via mixing followed by the Cabibbo-favoured (CF) decay \decay{\Dzb}{\Kp\pim} or via the direct decay \decay{\Dz}{\Kp\pim}, which is doubly Cabibbo-suppressed (DCS).
These two amplitudes interfere and the combined process is referred to as wrong-sign (WS) decay (see Fig.~\ref{fig:mix_dicovery}).
The corresponding Cabibbo-favoured process is calles right-sign (RS) decay.
This leads to the WS decay rate, using Eq.~\ref{eqn:m0t},
\begin{equation}
\Gamma(\decay{\Dz(t)}{\Kp\pim})=\left|\frac{q}{p}f_-(t)\bar{A}_{\bar{f}}+f_+(t)A_{\bar{f}}\right|^2,
\label{eqn:GammaWS}
\end{equation}
where $A_{\bar{f}}$ is the DCS amplitude of the \Dz decay and $\bar{A}_{\bar{f}}$ is the CF amplitude of the \Dzb decay.

Assuming \CP conservation, the time-dependent decay rate of the WS decay is
\begin{equation}
\Gamma(\Dz(t)\to \Kp\pim)=e^{-\Gamma t} \left(R_{\rm D}+\sqrt{R_{\rm D}}y'\Gamma{}t+R_{\rm m}(\Gamma{}t)^2\right),
\end{equation}
where $R_{\rm D}=|A_{\bar{f}}/\bar{A}_{\bar{f}}|^2$ is the ratio of the DCS to the CF rate.
The mixing parameters are rotated by the strong phase difference between the DCS and the CF amplitude, $\delta_{\PK\Ppi}$, leading to the observable $y'=y\cos\delta_{\PK\Ppi}-x\sin\delta_{\PK\Ppi}$~\cite{Bergmann:2000id}.

First evidence for charm mixing was obtained by the \babar collaboration in March 2007 based on a measurement of wrong-sign \decay{\Dz}{\Kp\pim} decays~\cite{Aubert:2007wf}.
While the two observables $x'^2$ and $y'$ were not measured to be non-zero with a significance exceeding three standard deviations, the no-mixing point $x'^2=y'=0$ is excluded with a significance of $3.9$ standard deviations.
Further supporting evidence was also obtained in the same measurement by the \cdf collaboration later in the same year~\cite{Aaltonen:2007uc}.
It then took until late 2012 for the \lhcb collaboration to perform a measurement excluding the no-mixing hypothesis by $9.1$ standard deviations, thus providing first single-measurement observation of mixing in the charm system~\cite{LHCb-PAPER-2012-038}.
This result can be seen in Fig.~\ref{fig:mix_dicovery} where the clearly non-zero slope indicates the presence of mixing.
Subsequent updates led to the following set of latest measurements

\begin{table}[h!!!]
\centering
\begin{tabular}{lcc}
           & $x'^2$ in $10^{-3}$ & $y'$ in $10^{-3}$\\
\hline
\cdf~\cite{Aaltonen:2013pja} & $0.08\pm0.18$ & $4.3\pm4.3$\\
\lhcb~\cite{LHCb-PAPER-2013-053} & $0.06\pm0.05$ & $4.8\pm1.0$\\
\belle~\cite{Ko:2014qvu} & $0.09\pm0.22$ & $4.6\pm3.4$
\end{tabular}
\end{table}

\noindent Similarly, the CF and DCS amplitudes can also lead to excited states of the same quark content.
The decay \decay{\Dz}{\Kp\pim\piz} is the final state of several such resonances, \eg\ $\Kp\rho^-$ and $\PK^{*+}\pim$.
Thus, by studying the decay-time dependence of the various resonances a mixing measurement can be obtained.
The \babar collaboration achieved a measurement showing evidence for \Dz mixing~\cite{Aubert:2008zh} with central values of $x''=(26.1^{+5.7}_{-6.8}\pm3.9)\times10^{-3}$ and $y''=(-0.6^{+5.5}_{-6.4}\pm3.4)\times10^{-3}$, where the rotation between the observables and the system of mixing parameters is given by a strong phase difference as 
\begin{align}
x''&=x\cos\delta_{\Kp\pim\piz}+y\sin\delta_{\Kp\pim\piz}\\*
y''&=y\cos\delta_{\Kp\pim\piz}-x\sin\delta_{\Kp\pim\piz}.
\end{align}
The significant advantage of this analysis over that using two-body final states is that both mixing parameters are measured at first order rather than one at first and one at second order.

The strong phase differences are not accessible in these measurements but have to come from measurements performed using quantum-correlated \Dz-\Dzb pairs produced at threshold.
Such measurements are available from \cleo~\cite{PhysRevD.73.034024,PhysRevD.77.019901,Rosner:2008fq,Asner:2008ft} and \besiii~\cite{Ablikim:2014gvw}.
In addition, further constraints on these strong phase differences can be obtained from the combination of several measurements that share the underlying mixing parameters but are subject to different strong phase differences.
This can be for example the combination of \decay{\Dz}{\Kp\pim} and \decay{\Dz}{\Kp\pim\piz} together with other measurements that are not affected by strong phases (see below).

{~\\ \noindent \bf Mixing in \CP eigenstates: }
In the absence of \CP violation the physical eigenstates are \CP eigenstates and therefore the width difference can be accessed directly through a measurement of the effective lifetime of a \CP eigenstate with respect to the lifetime of a flavour-specific state, which is not affected by mixing, using the observable
\begin{equation}
\ycp\equiv\frac{\Gamma_{\CP\pm}}{\Gamma}-1=\frac{\Gamma_{2,1}}{\Gamma}-1=\frac{\Delta\Gamma}{2\Gamma}\equiv y.
\end{equation}
In a more general scenario allowing also for \CP violation the effective lifetimes of \Dz (\Dzb) mesons into final states that are \CP eigenstates, $\hat{\tau}$ ($\hat{\bar{\tau}}$), lead to
\begin{equation}
\ycp=\frac{2\tau_{\PD}}{\hat{\tau}+\hat{\bar{\tau}}}-1\approx\eta_{\CP}\left[\left( 1 -\frac{a_{\rm m}^2}{8}\right)y\cos\phi -\frac{a_{\rm m}}{2}x\sin\phi\right],
\label{eqn:ycp}
\end{equation}
where $\tau_{\PD}$ is the \Dz lifetime and
\begin{equation}
\pm a_{\rm m}\equiv\frac{\left|\frac{q}{p}\right|^{\pm 2} - 1}{\left|\frac{q}{p}\right|^{\pm 2} + 1},
\end{equation}
which leads to $|q/p|^{\pm2}\approx 1\pm a_{\rm m}$~\cite{Gersabeck:2011xj}.
As the \CP-violating contributions $a_m$ and $\phi$ enter only at second order, measurements of \ycp are among the most powerful constraints of the mixing parameter $y$.

Following only a few weeks after the \babar measurement using \decay{\Dz}{\Kp\pim} decays, a measurement of the \belle collaboration of the parameter \ycp was among the first to provide evidence for mixing in the charm system~\cite{Staric:2007dt}.
The latest results are
\clearpage
\begin{table}[h!!!]
\centering
\begin{tabular}{lc}
           & \ycp in $10^{-3}$\\
\hline
\lhcb~\cite{Aaij:2011ad} & $5.5\pm6.3\pm4.1$\\
\babar~\cite{Lees:2012qh} & $7.2\pm1.8\pm1.2$\\
\belle~\cite{Staric:2012ta} & $11.1\pm2.2\pm1.1$
\end{tabular}
\end{table}

The \babar collaboration has added the larger sample of untagged events to their analysis; however, with limited gain in sensitivity due to larger systematic uncertainties for the untagged sample which has lower purity compared to the \Dstar-tagged events.
The \lhcb result is based on only about $1\%$ of their total dataset and significant improvements in precision are expected from future updates.

The central values of these results yield an average value of \ycp, which is greater than that of the mixing parameter $y$.
\CP violation can only mildly increase \ycp over $y$ (see Eq.~\ref{eqn:ycp}) and thus such a tension would not easily be explained.
However, based on the latest results, this discrepancy is at the level of two standard deviations even for the \belle result and thus considered not significant~\cite{Gersabeck:2014ega}.

Another possibility of measuring \ycp is using the decay mode \decay{\Dz}{\KS\Km\Kp}.
The \belle collaboration have published a measurement in which they compare the effective lifetime around the \Pphi resonance with that measured in sidebands of the $\Km\Kp$ invariant mass~\cite{Zupanc:2009sy}.
The effective \CP eigenstate content in these regions is determined with two different models.
Their result is $\ycp=(1.1\pm6.1\pm5.2)\times10^{-3}$.

Similarly, one can measure \ycp in a phase-space integrated way provided that the decay is close to being a \CP eigenstate~\cite{Malde:2015xra}.
Based on measurements of the CLEO collaboration the decay \decay{\Dz}{\pim\pip\piz} has been confirmed to be practically \CP even~\cite{Nayak:2014tea}.
Other potential candidates are  \decay{\Dz}{\pim\pip\pim\pip} and \decay{\Dz}{\KS\pim\pip\piz}, for which the effective \CP content remains to be measured.

{~\\ \noindent \bf Direct measurements of the mixing observables: }
\label{sec:mixing_KShh}
The decay \decay{\Dz}{\KS\Km\Kp} and more so the decay \decay{\Dz}{\KS\pim\pip} give excellent access to the mixing parameters $x$ and $y$ individually.
This is achieved through the simultaneous measurement of the decay-time evolution and resonance amplitudes in the Dalitz plot.
At the same time measurements of these final states allow a determination of parameters of indirect \CP violation as discussed in the following section.
Under the assumption of no \CP violation \belle and \babar have measured
\begin{table}[h!!!]
\centering
\begin{tabular}{lcc}
           & $x$ in $10^{-3}$ & $y$ in $10^{-3}$\\
\hline
\babar~\cite{delAmoSanchez:2010xz} & $1.6 \pm 2.3 \pm 1.2\pm0.8$ & $5.7 \pm 2.0 \pm 1.3\pm0.7$\\
\belle~\cite{Peng:2014oda} & $5.6 \pm 1.9 ^{+0.3+0.6}_{-0.9-0.9}$ & $3.0 \pm 1.5 ^{+0.4+0.3}_{-0.5-0.6}$
\end{tabular}
\end{table}

\noindent where the last uncertainty in each measurement is a model uncertainty.
With this analysis being arguably the most complicated charm analysis, there is no result from the \lhcb collaboration to date.
However, their large dataset should enable significant improvements of this measurement, which will be of particular benefit to contraining the mixing parameter $x$.

\clearpage
{~\\ \noindent \bf Combined results: }
As discussed above, no single measurement has thus far been able to pin down both mixing parameters $x$ and $y$.
Therefore, all available measurements are being combined by the Heavy Flavor Averaging group in order to obtain maximal sensitivity to the underlying parameters.
In doing so, measurements of strong phase differences are used, but also the combination itself is overconstraining and thus helps to pin down these phases.

Under the assumption of no \CP violation in mixing or decays, the world average of the mixing parameters is $x=(4.9^{+1.4}_{-1.5})\times10^{-3}$ and $y=(6.2\pm0.8)\times10^{-3}$~\cite{Amhis:2014hma}.
This result affirms the conclusion that the \CP even final state is the shorter-lived.
However, the periodic oscillation of \Dz and \Dzb mesons and vice versa, which is governed by the parameter $x$, has not yet been established.
The no-mixing hypothesis, \ie\ $R_{\rm m}=0$, is excluded by well over $12$ standard deviations.

\subsubsection{Measurement techniques}
\noindent Although all measurements discussed above are based on the decay-time evolution of \Dz decays, not all of the have been performed by a fit to an exponential distribution or modulation thereof.
Essentiall all measurements rely on the comparison of the decay-time evolution of two related processes.
The measurement based on semi-leptonic decays can be performed time-integrated as its time dependence is given by a single term proportional to $t^2$.

The measurement based on wrong-sign \decay{\Dz}{\Kp\pim} decays and that of \ycp is based on the comparison of the time evolution of two related decay modes.
This can be done by fitting the appropriate model, an exponential multiplied by a second order polynomial for the WS \Kp\pim decay or a simple exponential for \CP eigenstates, and comparing that to the lifetime measured in the CF \decay{\Dz}{\Km\pip} decay.
An alternative is to divide the data into bins of decay time and to measure the ratio of signal yields in these bins.
This ratio can then be fitted using a second order polynomial for the WS \Kp\pim decay or a linear function to extract \ycp.

Both techniques have been used in the different measurements and sometimes they were taken to provide mutual cross-checks in the same measurement.
A major advantage of the binned ratio method is that it cancels decay-time acceptance effects when they can be assumed to be sufficiently similar for the two modes being compared.
This is generally the case for the WS \Kp\pim decay measurement, but not for \ycp as the latter compared decays with different final state particles.
The decay-time acceptance argument is particularly relevant to hadron colliders where signal candidates are usually required to originate from some distance away from the primary collision point.
At \en\ep colliders the \Dz mesons are generally less boosted and therefore the decay-time resolution is worse.
In this context bin migration becomes an important factor to take into account, while the resolution is readily taken into account in an exponential fit.

The situation is more complicated for multi-body final states.
These generally require an analysis of the decay-time dependence of the different contributions to the phase space.
This can be obtained by two different groups of methods.
The first uses a model of the phase space and fits a dedicated model to the decay-time distribution each of the resonances or groups of resonances contributing to the phase-space model.
The decay-time models incorporate the common underlying theory parameters depending on the amplitudes contributing to the specific resonance.

The alternative method is based on measuring the decay-time dependence in specific regions of phase space.
These regions were chosen to enclose areas of similar strong phase difference and the choice of these regions is inspired by a phase-space model.
The effective \CP content of each of these regions needs to be known, which can be obtained through measurements of quantum-entangled \Dz mesons, such as in Ref.~\cite{Libby:2010nu}.
Using this information, the decay-time distributions can be linked to the underlying theory parameters once more.
The challenge of the former approach lies in assigning an uncertainty to the accuracy of the phase-space model, while the latter depends on accurate measurements of the \CP content of these regions.
While the accuracy of models may improve in the future through a combined effort of experiment and theory, any improvement on the second approach currently relies on further measurements made by the \besiii collaboration.

\section{Charm \CP violation}
\label{sec:cp_charm}

\subsection{Indirect \CP violation}
\subsubsection{Measuring indirect \CP violation directly} 
\noindent Indirect \CP violation is the deviation from unity of $q/p$, which comes with the fact that the mass eigenstates $|\PD_1\rangle$ and $|\PD_2\rangle$ are no longer \CP eigenstates.
This is measured most straight-forwardly by comparing the decay-time dependence of \Dz and \Dzb decays to \CP eigenstates.
Writing the effective lifetimes of the \Dz and \Dzb decays as $\hat{\tau}$ and $\hat{\bar{\tau}}$, respectively, this leads to the observable
\begin{equation}
\agamma=\frac{\hat{\bar{\tau}}-\hat{\tau}}{\hat{\bar{\tau}}+\hat{\tau}}\approx\eta_{\CP}\left[\frac{1}{2}\left( a_{\rm m}+a_{\rm d}\right)y\cos\phi -x\sin\phi\right]\approx -a_{\CP}^{\rm ind}+\frac{1}{2}a_{\rm d}\ycp,
\label{eqn:agamma}
\end{equation}
which has contributions from both direct and indirect \CP violation~\cite{Gersabeck:2011xj,Kagan:2009gb}.

All available measurements of \agamma to date have been based on the two-body \CP eigenstates $\Km\Kp$ and $\pim\pip$.
The \babar and \belle collaborations have published measurements based on the combination of these final states.
They obtain $\agamma=(0.9\pm2.6\pm0.6)\times10^{-3}$~\cite{Lees:2012qh} and $\agamma=(-0.3\pm2.0\pm0.8)\times10^{-3}$~\cite{Staric:2012ta}, respectively.
The \lhcb collaboration has measured \agamma separately for the two decay modes and separately for two flavour-tagging approaches.
The results based on a sample where \decay{\Dstarp}{\Dz\pip} decays are used for tagging the \Dz flavour are $\agamma(\Km\Kp)=(-0.4\pm0.6\pm0.1)\times10^{-3}$ and $\agamma(\Km\Kp)=(0.3\pm1.1\pm0.1)\times10^{-3}$~\cite{LHCb-PAPER-2013-054}, representing the most precise measurement of a charm \CP asymmetry to date.
As this result is based on only one third of \lhcb's integrated luminosity, further improvement may be expected with existing data.
In addition, a recent analysis is based on a complementary data sample where the \Dz flavour is tagged using the charge of the muon from semi-leptonic \bquark-hadron decays involving \Dz mesons.
The results are $\agamma(\Km\Kp)=(-1.3\pm0.8\pm0.3)\times10^{-3}$ and $\agamma(\Km\Kp)=(-0.9\pm1.5\pm0.3)\times10^{-3}$~\cite{Aaij:2015yda}.
All individual measurements are in good agreement with \CP conservation.
A naive average assuming negligible final state differences yields $\agamma=(-0.6\pm0.4)\times10^{-3}$.

Using current experimental bounds values of \agamma up to $\mathcal{O}(10^{-4})$ are expected from theory~\cite{Kagan:2009gb,Bigi:2011re}.
It has however been shown that enhancements up to about one order of magnitude are possible, for example in the presence of a fourth generation of quarks~\cite{Bobrowski:2010xg} or in a little Higgs model with T-parity~\cite{Bigi:2011re}. 
The current level of precision starts to put bounds on this parameter space and can be expected to be further improved with the analysis of the full LHC run-1 dataset of the \lhcb experiment.

Eventually, the interpretation of the observable \agamma requires precise knowledge of both mixing and \CP violation parameters.
The relative sensitivity to the \CP-violating quantities in \agamma is limited by the relative uncertainty of the mixing parameters.
Therefore, to establish the nature of a potential non-zero measurement of \agamma it is mandatory to have measured the mixing parameters with a relative precision of about $10\%$.

\subsubsection{\CP asymmetries in \decay{\Dz}{\Kp\pim} decays}
\label{sec:cpv_wskpi}
\noindent Further powerful access to \CP asymmetries lies in \decay{\Dz}{\Kp\pim} decays.
When conducting the measurement described in Sec.~\ref{sec:mixing_measurements} separately for \Dz and \Dzb decays one can define the set of observables $R_{\rm D}^\pm$, $x'^{2\pm}$, and $y'^\pm$, where the $\pm$ refers to the parameter for the \Dz and \Dzb decay, respectively.
Following from Eq.~\ref{eqn:GammaWS}, these are related to the underlying parameters by
\begin{eqnarray}
x'^\pm & = & \left|\frac{q}{p}\right|^{\pm1}(x'\cos\phi\pm y'\sin\phi)\nonumber\\
y'^\pm & = & \left|\frac{q}{p}\right|^{\pm1}(y'\cos\phi\mp x'\sin\phi),
\end{eqnarray}
and a non-zero difference between $R_{\rm D}^+$ and $R_{\rm D}^-$ would indicate \CP violation in the DCS decay.
For small phases $\phi$ this gives direct access to the magnitude of $|q/p|$ and thus \CP violation in mixing.
For larger weak phases the sensitivity is worse since the sensitivity to $x'$ is much worse compared to $y'$.
Several measurements of these parameters have been made by \belle~\cite{Zhang:2006dp}, \babar~\cite{Aubert:2007wf}, and \lhcb~\cite{LHCb-PAPER-2013-053} and none of them shows any indication of \CP violation.

\subsubsection{Direct measurements of $|q/p|$ and $\phi$}
\noindent The analyses of the decays \decay{\Dz}{\KS\pim\pip} and \decay{\Dz}{\KS\Km\Kp} offer separate access to the parameters $x$, $y$, $|q/p|$ and $\arg(q/p)$ and are one of the most promising ways of obtaining precise mixing measurements.
These analyses require the determination of the decay-time dependence of the phase space structure as described for the mixing measurements in Sec.~\ref{sec:mixing_KShh}.
The only existing measurement to include the \CP violating parameters has been performed by the \belle collaboration to yield $|q/p|=0.90^{+0.16+0.05+0.06}_{-0.15-0.04-0.05}$ and $\phi=(-6\pm11\pm3^{+3}_{-4})^\circ$~\cite{Peng:2014oda}.
Other measurements were performed by the \cleo~\cite{Asner:2005sz} and \babar~\cite{delAmoSanchez:2010xz} collaborations assuming \CP conservation and thus extracting only $x$ and $y$.
With the data samples available and being recorded at \lhcb and those expected at future flavour factories, these measurements will be very important to understand charm mixing and \CP violation.
However, in order to avoid systematic limitations it will be important to reduce model uncertainties or to improve model-independent strong-phase difference measurements which are possible at \besiii.

\begin{figure}[tbh]
\centering
\includegraphics[width=0.7\textwidth]{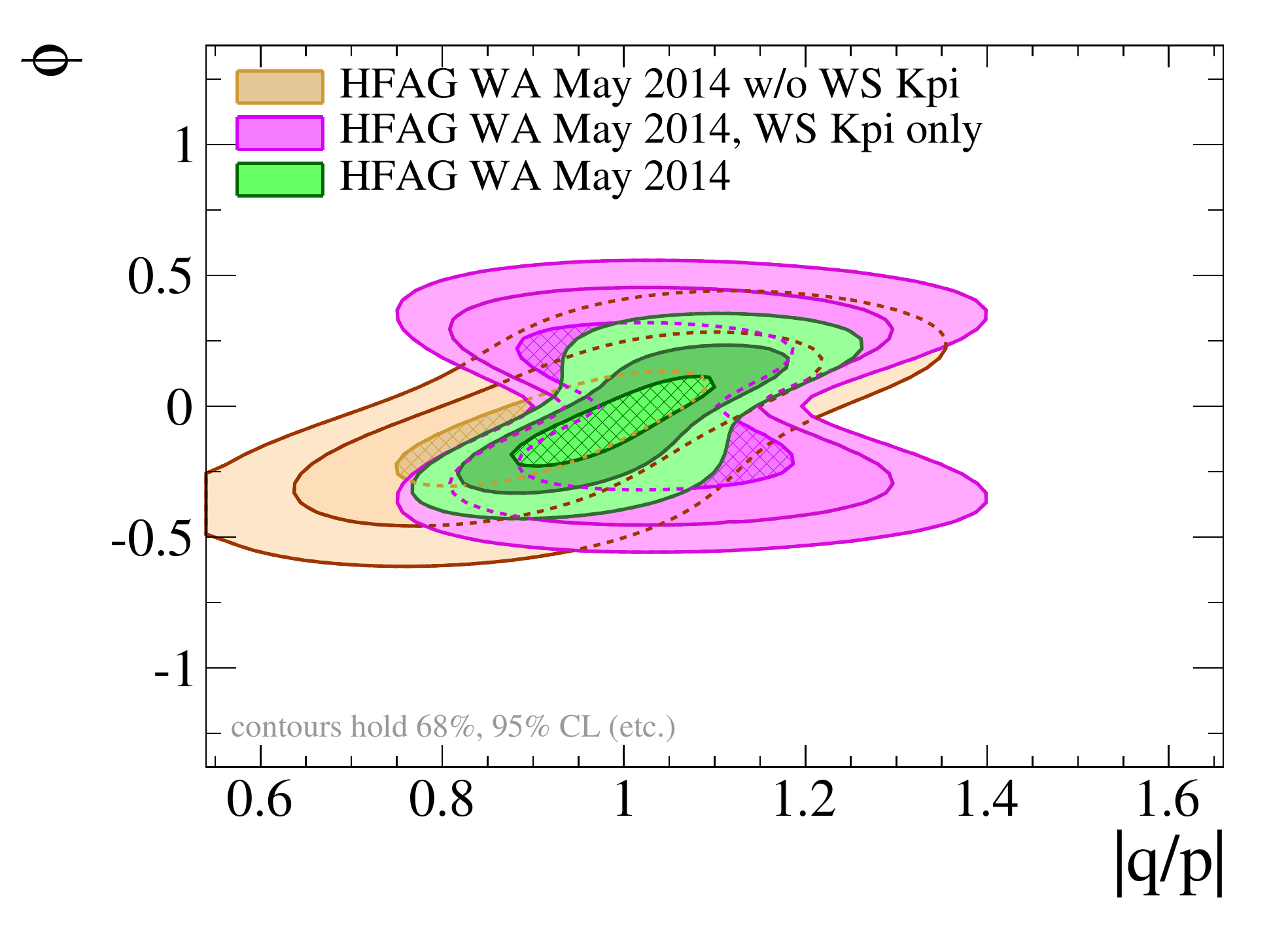}
\caption{Contributions of the combination of different measurements to the extraction of the parameters $|q/p|$ and $\phi$. Reproduced from Ref.~\cite{Gersabeck:2014ega}.}
\label{fig:qoverp_phi}
\end{figure}

\subsubsection{Combining indirect \CP violation measurements}
\noindent Direct access to \CP violation parameters is limited to the one class of measurements described in the previous paragraph.
Therefore, extracting the underlying theory parameters from a combined fit to all relevant results is required.
Figure~\ref{fig:qoverp_phi} shows how different measurements contribute to the current world average as computed by the Heavy Flavor Averaging Group.
The brown contour in the background essentially combines measurements of \decay{\Dz}{\KS\pim\pip}, which lead to an elliptical constraint, and measurements of \agamma, which constrain a diagonal in this parameter space.
The magenta contour shows the power of measurements based on \decay{\Dz}{\Kp\pim} decays alone, underlining the particular sensitivity to $|q/p|$ for small values of $\phi$.
Finally, the green contour in the foreground is the full combination of all measurements.
The official world average is $|q/p|=0.93^{+0.09}_{-0.08}$ and $\phi=(-8.7^{+8.7}_{-9.1})^\circ$~\cite{Amhis:2014hma}.

Under the assumption that there is no \CP violation in the corresponding decay rates one can construct the following relationship~\cite{Ciuchini:2007cw,Kagan:2009gb,Grossman:2009mn}:
\begin{equation}
\tan\phi=\frac{1-|q/p|^2}{1+|q/p|^2}\times\frac{x}{y}.
\end{equation}
This equation can be used to reduce the number of mixing and indirect \CP violation parameters from four to three.
Using this to remove $\phi$ the world average on $|q/p|$ becomes $1.007^{+0.015}_{-0.014}$ and correspondingly, removing $|q/p|$ one obtains $\phi=(-0.30^{+0.58}_{-0.60})^\circ$~\cite{Amhis:2014hma}, which shows an impressive improvement in sensitivity.

\subsection{Direct \CP violation}
\noindent Direct \CP violation is the asymmetry of a decay rate to its \CP conjugate, defined as
\begin{equation}
a_{\CP}(\decay{\PD}{f})\equiv\frac{\Gamma(\decay{\PD}{f})-\Gamma(\decay{\Dbar}{\bar{f}})}{\Gamma(\decay{\PD}{f})+\Gamma(\decay{\Dbar}{\bar{f}})},
\end{equation}
for the decay of a \PD meson to a final state $f$.
Direct \CP asymmetries are generally measured in decay-time integrated measurements.
For neutral mesons, the decay-time distribution of the data has to be taken into account to estimate the contribution from indirect \CP violation as will be further explained in Sec.~\ref{sec:cpv_interplay}.

\subsubsection{Experimental challenges}
\noindent Experimentally, there are a number of effects that can generate asymmetries and thus obscure access to the physics asymmetry $a_{\CP}$.
Under the assumption of small asymmetries, these effects are additive to first order and can be written as
\begin{equation}
a_{\rm raw} = a_{\CP} + a_{\rm det,f} + a_{\rm det,tag} + a_{\rm prod},
\end{equation}
where $a_{\rm det,f}$ and $a_{\rm det,tag}$ are detection asymmetries of the final state and the tagging process, respectively, and where $a_{\rm prod}$ is the production asymmetry of the initial state.

The production asymmetry is only an issue at matter anti-matter asymmetric machines.
These are $pp$ colliders like the LHC or fixed target experiments where a matter beam is shot on a matter target.
However, $p\bar{p}$ colliders such as the Tevatron can have a forward-backward asymmetry as protons and anti-protons have a preferred direction.
A forward-backward asymmetry exists also at $\ep\en$ colliders where its origin lies in the interference of particle production via virtual photons and via virtual $Z$ bosons.
Experimentally, production asymmetries are measured using control modes with the same initial state as the signal decay.

The detection asymmetries have two origins.
These are detector asymmetries where the detector geometry, mostly in combination with a magnetic field, leads to a higher probability of detecting particles from one charge compared to the other.
One example can be caused by staggering of detector elements, which can lead to an acceptance gap for particles travelling in a specific direction while it is fine for particles traversing the detector at a different angle.
The second component are interaction asymmetries, which are based on the fact that particles of one charge have a different probability of interacting with the detector than the particle of the other charge.
One example are \Km mesons for which the strange quark can produce hyperons and the anti-up quark can annihilate with the detector material, whereas the anti-strange quark of the \Kp can neither produce hyperons or annihilate with quarks in the detector material.
Detection asymmetries may intrinsically cancel if two particles of the same kind and with opposite charge are present in the final state, such as in the decay \decay{\Dz}{\Km\Kp}.
Otherwise, detection asymmetries need to be measured in control modes and subtracted off.

Due to the underlying mechanisms that cause production and detection asymmetries they may vary with the particle's kinematics or its position in the detector.
Therefore, any subtraction of asymmetries needs to ensure sufficient overlap in the relevant quantities to guarantee proper cancellation.

\subsubsection{Experimental measurements}
\label{sec:cpv_direct_meas}
{\noindent \bf \CP violation in two-body final states: }
Keeping in mind the above, a very powerful observable is the difference in \CP asymmetries of the decays of \Dz mesons into two charged pions or kaons.
In this case all nuisance asymmetries cancel to first order and the measured difference is equal to the difference in physics observables, defined as
\begin{displaymath}
\dacp\equiv{}a_{\CP}(\Km\Kp)-a_{\CP}(\pim\pip)=a_{\rm raw}(\Km\Kp)-a_{\rm raw}(\pim\pip).
\end{displaymath}

An initial measurement by the \lhcb collaboration was the origin of an increase in experimental and particularly theoretical activity in this field.
The result of $\dacp=(-8.1\pm2.1\pm1.1)\times10^{-3}$~\cite{Aaij:2011in} provided first evidence for \CP violation in the charm sector.
Subsequent measurements did thus far not confirm this evidence and the latest world average of $(-2.5\pm1.0)\times10^{-3}$ already excludes percent level \CP violation.
Measurements from \babar~\cite{Aubert:2007if}, \belle~\cite{Ko:2012px}, \cdf~\cite{Collaboration:2012qw}, and \lhcb~\cite{LHCb-CONF-2013-003} are based on \Dstarp-tagged decays, while the \lhcb collaboration has also published a measurement based on \Dz mesons originating in semi-leptonic \bquark-hadron decays~\cite{Aaij:2014gsa}.

The observable \dacp gives access to the difference in direct \CP violation of the two decay modes through
\begin{equation}
\dacp=\Delta{}a_{\CP}^{\rm dir}\left(1+\ycp\frac{\overline{\langle{}t\rangle}}{\tau}\right)-\overline{A}_\Gamma\frac{\Delta\langle{}t\rangle}{\tau},
\label{eqn:dacp}
\end{equation}
where $\tau$ is the nominal \Dz lifetime, $\overline{X}\equiv(X(\Km\Kp)+X(\pim\pip))/2$, and $\Delta{}X\equiv{}X(\Km\Kp)-X(\pim\pip)$ for $X=(a_{\CP}^{\rm dir},\langle{}t\rangle,\agamma)$~\cite{Gersabeck:2011xj}.
Here, $\overline{A}_\Gamma$ is multiplied with the difference of mean decay times as indirect \CP violation leads to different effective lifetimes.
Thus, its impact on a time-integrated asymmetry depends on the decay-time distribution of the measured data sample.
Equation~(\ref{eqn:dacp}) assumes the \CP-violating phase $\phi$ to be universal.
For a small non-zero difference in this phase between the two final states, $\Delta\phi_{f}$, an additional term of the form $x\Delta\phi_{f}\overline{\langle{}t\rangle}/\tau$ arises as pointed out in Ref.~\cite{Kagan:2009gb}.
Given a typical variation of $\overline{\langle{}t\rangle}/\tau$ between $1$ and $2.5$ for the different experiments the contribution of $\Delta\phi_{f}$ is suppressed by $x\overline{\langle{}t\rangle}/\tau\approx10^{-2}$.

An additional beneficial feature of the observable \dacp is that the two contributing asymmetries are expected to be roughly of equal magnitude and opposite sign, due to the CKM structure of the decays.
Thus, \dacp would measure twice the magnitude of one of the individual asymmetries.
Nevertheless, in addition to measuring the difference, the knowledge of the individual asymmetries is required to pin down the source of potential \CP violation.
Several measurements of the individual asymmetries exist, with the most recent and most precise one by the \lhcb collaboration~\cite{Aaij:2014gsa}.

The challenge in measuring these individual asymmetries clearly lies in controlling all nuisance asymmetries to extract the $a_{\CP}$ from the measured asymmetry.
The \lhcb measurement is based on flavour-tagging through semi-leptonic \bquark-hadron decays and their production asymmetry is controlled through a measurement of Cabibbo-favoured \decay{\Dz}{\Km\pip} decays produced in the same way.
While this also cancels the detection asymmetry of the tagging muon, it introduces a detection asymmetry due to the $\Km\pip$ final state.
This in turn is measured in the difference of the CF decays \decay{\Dp}{\Kp\pim\pim} and \decay{\Dp}{\KS\pim}, where in the latter decay the \KS decays into $\pip\pim$.
This leaves the asymmetry introduced by the \KS due to regeneration in the detector material and due to \CP violation in kaons.
Both effects are well known and can be calculated and subtracted to sufficient precision.
The latest world average is compatible with both asymmetries being of equal magnitude and opposite sign with $a_{\pi\pi}=(1.4\pm1.5)\times10^{-3}$ and $a_{\pi\pi}=(-1.1^{+1.4}_{-1.3})\times10^{-3}$~\cite{Amhis:2014hma}.

{~\\ \noindent \bf \CP violation in two-body final states of charged \PD mesons: }
Decays of \Dp and \Ds into a \KS and either a \Kp or a \pip are closely related to their \Dz counterparts.
Measurements of time-integrated asymmetries in these decays are expected to exhibit a contribution from \CP violation in the kaon system.
As has been pointed out~\cite{Grossman:2011to} this contribution depends on the decay-time acceptance of the \KS.
This can lead to different expected values for different experiments.
For \belle~\cite{Ko:2012pe}, the expected level of asymmetry due to \CP violation in $\Kz-\Kzb$ mixing is $-3.5\times10^{-3}$.
For \lhcb on the other hand, there is no significant asymmetry induced by kaon \CP violation~\cite{LHCb:2012fb} as the \lhcb acceptance, for \KS reconstructed in the vertex detector, corresponds to about $10\%$ of a \KS lifetime.
\CP violation searches in the decays \decay{\Dp}{\KS\pip}~\cite{PhysRevD.80.071103,Mendez:2009aa,PhysRevLett.104.181602} and \decay{\Ds}{\KS\pip}~\cite{Mendez:2009aa,PhysRevLett.104.181602} show significant asymmetries.
However, these asymmetries are fully accommodated in the expected \CP violation of the kaon system.
These measurements do not show any hint for an asymmetry in \PD decay amplitudes.
This is confirmed by recent \lhcb measurements in which kaon \CP violation is negligible~\cite{Aaij:2014qec}.

{~\\ \noindent \bf \CP violation in multi-body final states: }
Another group of channels suitable for \CP violation searches is that of decays of \Dp and \Ds mesons into three charged hadrons, namely pions or kaons.
Here, \CP violation can occur in quasi two-body resonances contributing to these decay amplitudes.
Asymmetries in the Dalitz-plot substructure can be measured using an amplitude model or using model-independent statistical analyses~\cite{Aubert:2008yd,Bediaga:2009tr,Aaij:2013jxa,Williams:2011cd,Aaij:2014afa}.
The latter methods allow \CP violation to be discovered while eventually a model-dependent analysis is required to identify its source.
The two types of model-independent analyses differ in being either binned~\cite{Aubert:2008yd,Bediaga:2009tr,Aaij:2013jxa} or unbinned~\cite{Aaij:2013jxa,Williams:2011cd,Aaij:2014afa} in the Dalitz plane.

\begin{figure}[tbh]
\centering
\includegraphics[width=0.49\textwidth]{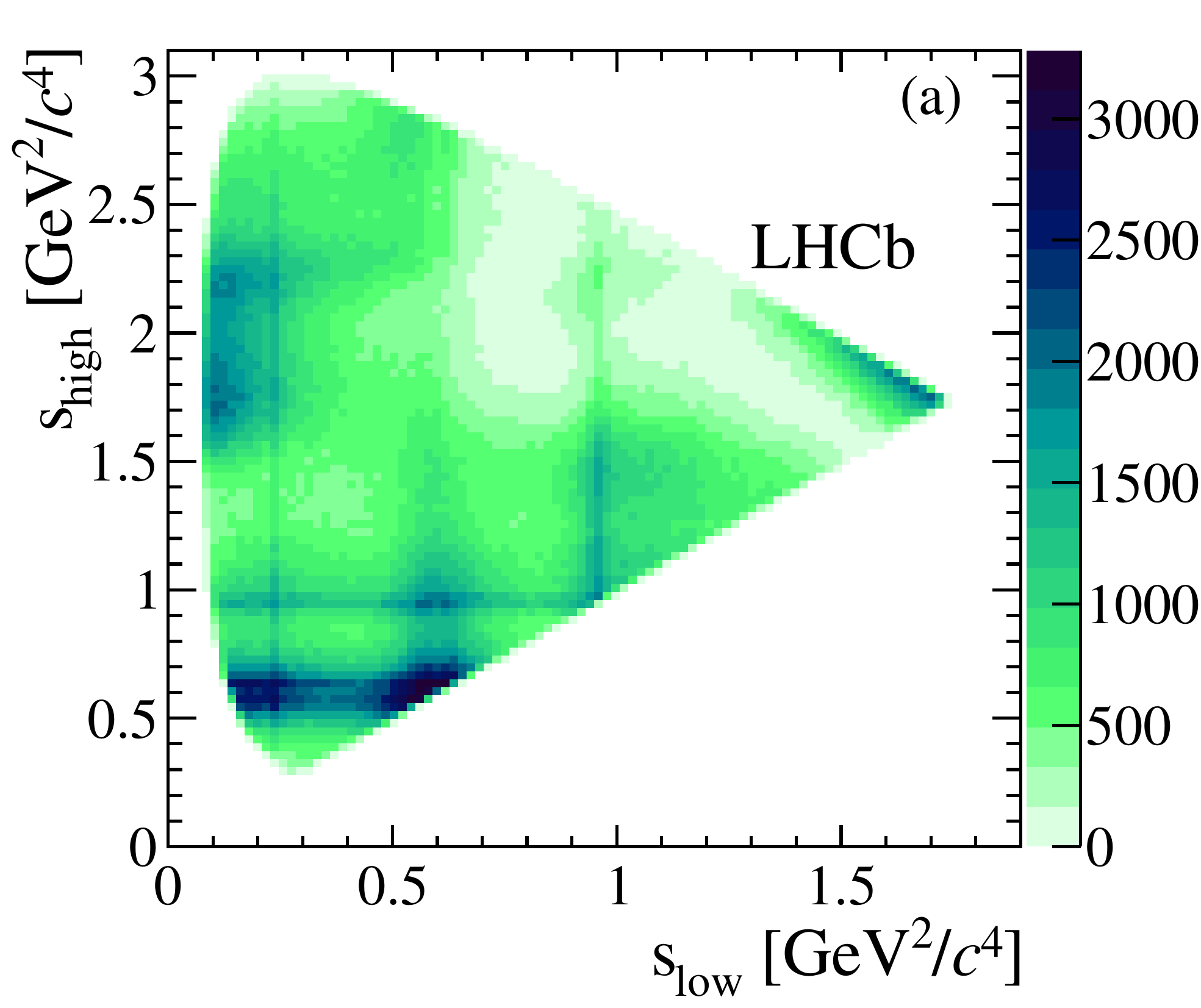}
\includegraphics[width=0.49\textwidth]{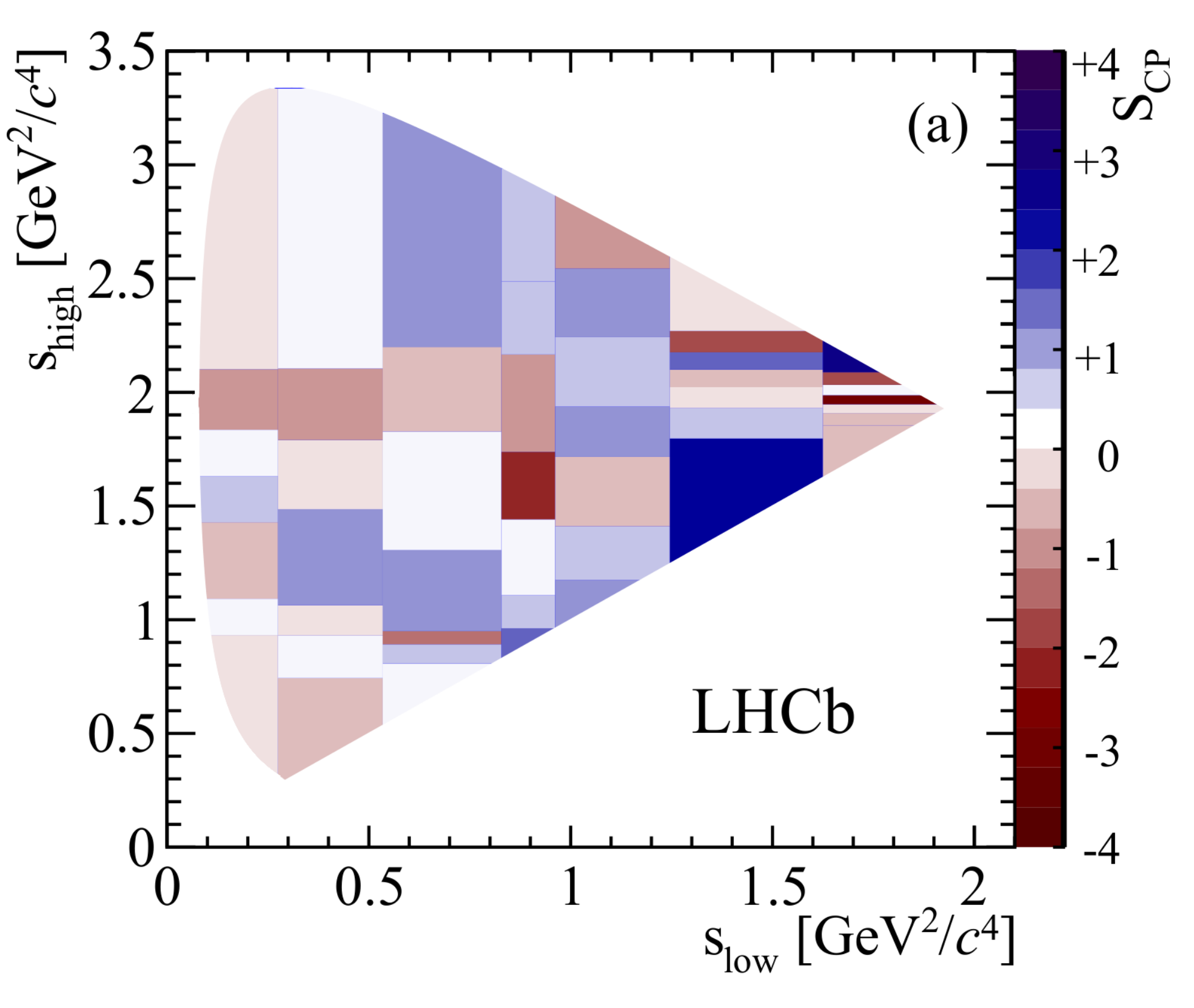}
\caption{(Left) Dalitz plot of \decay{\Dp}{\pip\pim\pip} candidates and (right) asymmetry significance $S_{\CP}$ for the same candidates. Reproduced from Ref.~\cite{Aaij:2013jxa}.}
\label{fig:cpv_scp}
\end{figure}

The binned approach computes a local per-bin asymmetry significance, $S_{\CP}$, and judges the presence of \CP violation by the compatibility of the distribution of local asymmetries across the Dalitz plane with a normal distribution.
An example of a Dalitz plot and corresponding $S_{\CP}$ distribution is shown in Fig.~\ref{fig:cpv_scp}.
This method obviously relies on the optimal choice of bins.
Bins ranging across resonances can lead to the cancellation of real asymmetries within a bin.
Too fine binning can reach the limit of statistical sensitivity, whereas too coarse binning can wash out \CP violation effects by combining regions of opposite asymmetry.
A model-inspired choice of binning is clearly useful and this does not create a model-dependence in the way that fitting resonances directly does.
This method does not yield an easy-to-interpret quantitative result.
This issue has been discussed in a recent update of the procedure~\cite{Bediaga:2012tm}.

\begin{figure}[tbh]
\centering
\includegraphics[width=0.49\textwidth]{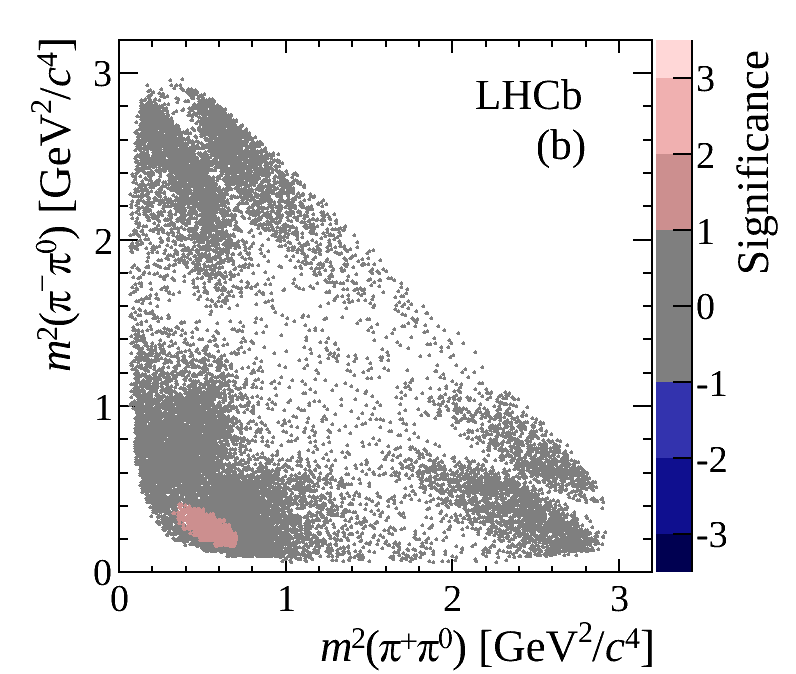}
\caption{Local asymmetry significances in the Dalitz plot of \decay{\Dz}{\pim\pip\piz} candidates. Reproduced from Ref.~\cite{Aaij:2014afa}.}
\label{fig:cpv_williams}
\end{figure}

The unbinned asymmetry search calculates a test statistic that allows the assignment of a $p$-value when comparing to the distribution of the statistic for many random permutations of the events among the particle and anti-particle datasets~\cite{Williams:2011cd}.
Moreover, being unbinned, there is no need for a model-inspired choice of binning.
The drawback of this method is its requirement on computing power.
The calculation of the test statistic scales as the square of the number of events.
In a first implementation of the method this challenge has been overcome by the usage of graphics processing units (GPUs)~\cite{Aaij:2014afa}.
Figure~\ref{fig:cpv_williams} shows the distribution of local asymmetry significances contributing to the above mentioned $p$-value calculation.
The unbinned technique exploited in Ref.~\cite{Aaij:2013jxa} can be seen as a simplification of the above procedure since it restricts the comparison to events that are nearest neighbours in phase space.
This comparison being applied to the order of 100 nearest neighbours for millions of signal candidates clearly reduces the computing requirement, but similarly reduces the method's sensitivity.

Beyond three-body final states similar analyses can be performed in decays into four hadrons, \eg\ decays of \Dz into four charged hadrons.
This too gives access to interesting resonance structures that may exhibit significant \CP asymmetries.
However, rather than having a two-dimensional Dalitz plane, the phase space for four-body decays is five-dimensional (see \eg\ Ref.~\cite{Rademacker:2006zx}).
This poses not only a challenge on the visualisation but also on any binned approach due to rapidly decreasing sample sizes per bin.
Also, the phase-space substructure can no longer be described only by interfering amplitudes of pseudo two-body decays as also three-body decays may contribute.
The \lhcb collaboration has released a first model-independent search for \CP violation in the decays \decay{\Dz}{\Km\Kp\pip\pim} and \decay{\Dz}{\pim\pip\pip\pim} without finding any hint of \CP non-conservation~\cite{Aaij:2013swa}.

{~\\ \noindent \bf Theory of charm \CP violation: }
The initially large central value for \dacp led to an intensive discussion on the potential origin of this \CP asymmetry.
In particular, it could not be ruled out that such a value could be caused by standard model effects.
On the other hand, since the size of standard model effects could not be pinned down precisely it could similarly not be ruled out that at least part of the effect was caused by physics beyond the standard model.
With the current world average being significantly smaller, it is more conceivable that an asymmetry of this size is in the range of standard model effects but other sources can still not be ruled out.

Within the standard model the central value can only be explained by significantly enhanced penguin amplitudes.
This enhancement is conceivable when estimating flavour $SU(3)$ or U-spin breaking effects from fits to data of \PD decays into two pseudo scalars~\cite{Feldmann:2012js,Bhattacharya:2012ah,Pirtskhalava:2011va,Brod:2012ud,Golden:1989qx,Cheng:2012xb}.
U-spin describes a subgroup of $SU(3)$ that exchanges \dquark-quarks and \squark-quarks.
However, attempts of estimating the long distance penguin contractions directly have failed to yield conclusive results to explain the enhancement.

Lattice QCD has the potential of assessing the penguin enhancement directly.
However, several challenges arise which make these calculations impossible at the moment~\cite{Luscher:1986pf,Luscher:1990ux,Lellouch:2000pv,Blum:2011ng,Blum:2011pu,Hansen:2012tf,Yu:2011gk}.
Following promising results on $\PK\to\pi\pi$ decays, additional hurdles arise in the charm sector as $\pi\pi$ and $\PK\PK$ states mix with $\eta\eta$, $4\pi$, $6\pi$ and other states.
Possible methods have been proposed and first results may be expected within the next decade.

General considerations on the possibility of interpreting \dacp in models beyond the standard model have lead to the conclusion that an enhanced chromomagnetic dipole operator is required.
These operators can be accommodated in minimal supersymmetric models with non-zero left-right up-type squark mixing contributions or, similarly, in warped extra dimensional models~\cite{Grossman:2006jg,Isidori:2011qw,Giudice:2012qq,Randall:1999ee,Goldberger:1999uk,Huber:2000ie,Gherghetta:2000qt}.
Tests of these interpretations beyond the standard model are in the focus of ongoing searches.
One promising group of channels are radiative charm decays where the link between the chromomagnetic and the electromagnetic dipole operator leads to predictions of enhanced \CP asymmetries of several percent~\cite{Isidori:2012yx}.

Another, complementary, test is to search for contributions beyond the standard model in $\Delta I=3/2$ amplitudes.
This class of amplitudes leads to several isospin relations which can be tested in a range of decay modes, e.g.\ $\PD\to\pi\pi$, $\PD\to\rho\pi$, $\PD\to\PK\bar{\PK}$~\cite{Feldmann:2012js,Grossman:2012eb}.
Several of these measurements, such as the Dalitz plot analysis of the decay $\Dz\to\pip\pim\piz$, have been performed by \babar and \belle and will be possible at \lhcb as well as future $\ep\en$ machines.

The \belle collaboration has recently published measurements of the challenging decays to neutral final states \decay{\Dz}{\piz\piz} and \decay{\Dz}{\KS\piz}.
They find $A(\piz\piz)=(-0.3\pm6.4\pm1.0)\times10^{-3}$ and $A(\KS\piz)=(-2.1\pm1.6\pm0.7)\times10^{-3}$~\cite{Nisar:2014fkc}.

Beyond charm physics, the chromomagnetic dipole operators would affect the neutron and nuclear electric dipole moments (EDMs), which are expected to be close to the current experimental bound~\cite{Giudice:2012qq}.
Similarly, rare flavour-changing neutral current (FCNC) top decays are expected to be enhanced.
Furthermore, quark compositness can be related to the $\Delta{}a_{\CP}$ measurement and tested in dijet searches.
Current results favour the new physics contribution to be located in the $\Dz\to\Km\Kp$ decay as the strange quark compositness scale is less well constrained~\cite{Bediaga:2012py}.

In the light of the recent measurements it is evident that there are four directions to pursue:
\begin{itemize}
\item More precise measurements of \dacp and the corresponding individual asymmetries are required to confirm the size of potential \CP violation in these decays.
\item Further searches for time-integrated \CP violation need to be carried out in a large range of decay modes that allow the identification of the source of the \CP asymmetry.
\item Searches for time-dependent \CP asymmetries, particularly via more precise measurements of \agamma, \decay{\Dz}{\Kp\pim} decays, and \decay{\Dz}{\KS\pim\pip} decays are required to establish or constrain the size of indirect \CP violation.
\item And finally a more precise determination of the mixing parameters, in particular $x$, is required to fully exploit the other observables.
\end{itemize}

\subsection{Interplay of mixing, direct and indirect \CP violation}
\label{sec:cpv_interplay}
\begin{figure}
\centering
\includegraphics[width=0.85\textwidth]{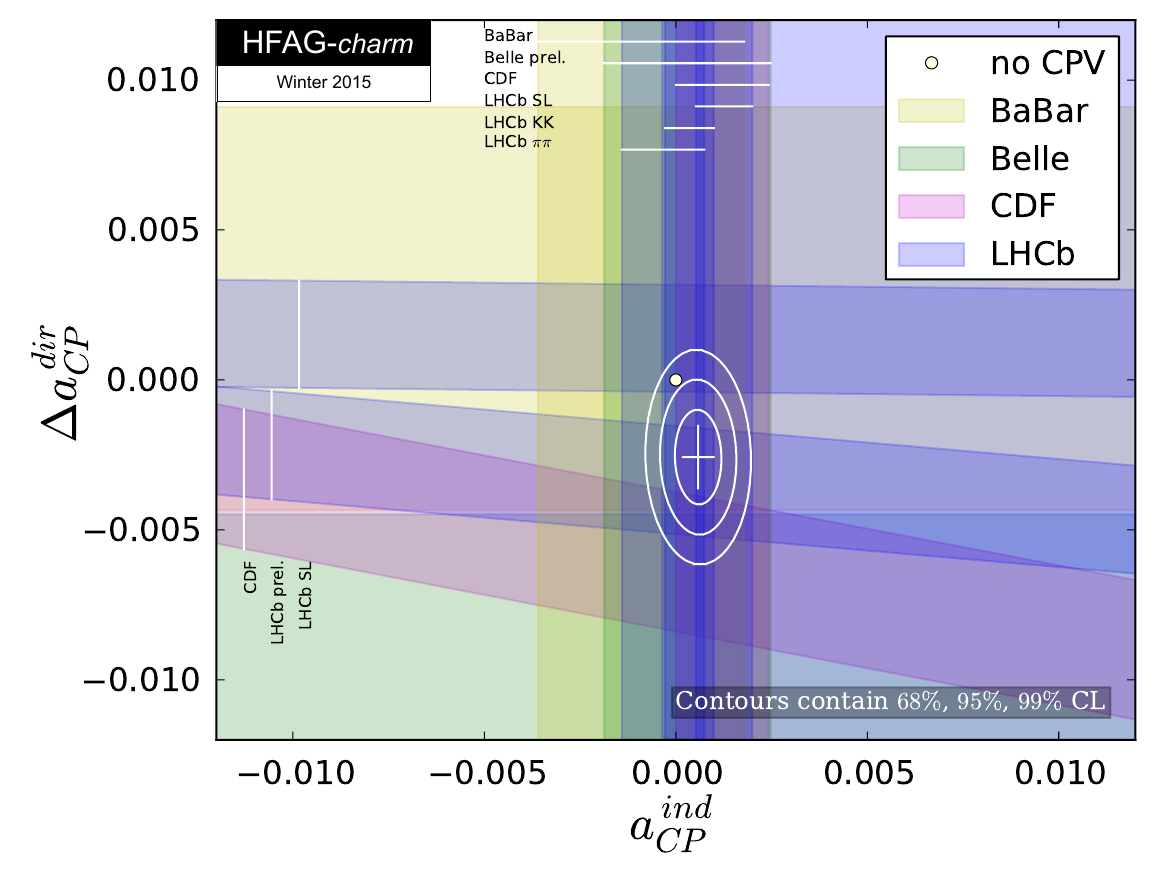}
\caption{Fit of $\Delta{}a_{\CP}^{\rm dir}$ and $a_{\CP}^{\rm ind}$. Reproduced from Ref.~\protect\cite{Amhis:2014hma}.}
\label{fig:hfag_cpv}
\end{figure}

\noindent Following Eqs.~(\ref{eqn:agamma}) and~(\ref{eqn:dacp}) it is obvious that both \agamma and \dacp share the underlying \CP-violating parameters.
Allowing for a non-universal \CP-violating phase $\phi$ one can write
\begin{align}
\agamma(f)&=-a_{\CP}^{\rm ind}-a_{\CP}^{\rm dir}(f)\ycp-x\phi_{f},\label{eq:agamma}\\
a_{\CP}(f)&=a_{\CP}^{\rm dir}(f)-\agamma(f)\frac{\langle{}t\rangle}{\tau},\\
\dacp&=\Delta{}a_{\CP}^{\rm dir}-\Delta\agamma\frac{\overline{\langle{}t\rangle}}{\tau}-\overline{A}_\Gamma\frac{\Delta\langle{}t\rangle}{\tau},
\end{align}
where again $\overline{X}\equiv(X(\Km\Kp)+X(\pim\pip))/2$ and $\Delta{}X\equiv{}X(\Km\Kp)-X(\pim\pip)$ for $X=(a_{\CP}^{\rm dir},\langle{}t\rangle,\agamma)$.
It is expected that, at least within the standard model, one has $a_{\CP}^{\rm dir}(\Km\Kp)=-a_{\CP}^{\rm dir}(\pim\pip)$ and thus $\overline{A}_\Gamma=-a_{\CP}^{\rm ind}$.
This set of equations shows that it is essential to measure both time-dependent (\agamma) and time-integrated asymmetries ($a_{\CP}$) separately in the decay modes \decay{\Dz}{\Km\Kp} and \decay{\Dz}{\pim\pip} in order to distinguish the various possible sources of \CP violation.
Currently, the experimental precision on \agamma is such that there is no sensitivity to differences in the contributions from direct \CP violation to measurements using $\Km\!\Kp$ or $\pim\!\pip$ final states.
Hence, the approximation $\agamma\equiv\overline{A}_\Gamma\approx\agamma(\Km\Kp)\approx\agamma(\pim\pip)$ can be used to obtain
\begin{align}
\agamma&=-a_{\CP}^{\rm ind}\\
\dacp&=\Delta{}a_{\CP}^{\rm dir}\left(1+\ycp\frac{\overline{\langle{}t\rangle}}{\tau}\right)+a_{\CP}^{\rm ind}\frac{\Delta\langle{}t\rangle}{\tau}.
\end{align}
These equations have been used by HFAG to prepare a fit of the direct and indirect \CP violation contributions~\cite{Amhis:2014hma} as shown in Fig.~\ref{fig:hfag_cpv}.
This fit yields a confidence level of about $1.8\times10^{-2}$ for the no \CP violation hypothesis with best fit values of $\Delta{}a_{\CP}^{\rm dir}=(-2.57\pm1.04)\times10^{-3}$ and $a_{\CP}^{\rm ind}=(0.58\pm0.40)\times10^{-3}$.
While there may be direct \CP violation at the $10^{-3}$ level, indirect \CP violation is constrained to be less than $10^{-3}$. 

The ultimate goal of mixing and \CP violation measurements in the charm sector is to reach precisions at or below the standard model predictions.
In some cases this requires measurements in several decay modes in order to distinguish enhanced contributions of higher order standard model diagrams from effects caused by new particles.

A dominant contribution to indirect \CP violation measurements is given by the observable \agamma (see Eq.~(\ref{eq:agamma})).
The \CP violating parameters in this observable are multiplied by the mixing parameters $x$ and $y$, respectively.
Hence, the relative precision on the \CP violating parameters is limited by the relative precision of the mixing parameters.
Therefore, aiming at a relative precision below $10\%$ for the \CP violation quantities and taking into account the current mixing parameter world averages, the target precision for the mixing parameters is $2-3\times10^{-4}$, corresponding to about $5\%$ of the current uncertainty of the world averages.
With standard model indirect \CP violation expected of the order of $10^{-4}$, the direct \CP violation parameter contributing to \agamma has to be measured to an absolute precision of $10^{-3}$ in order to distinguish the two types of \CP violation in \agamma.

Expectations for direct \CP violation vary for different decay modes.
In addition to the modes discussed so far, radiative decays have been pointed out as promising probes for physics beyond the standard model.
Possible \CP asymmetries exceeding $1\%$ are expected in \decay{\Dz}{V\gamma} ($V=\rho,\phi$) decays in the tails of the invariant mass distribution of the vector resonance~\cite{Isidori:2012yx}.
These regions have particularly high sensitivity to contributions through dipole operators.

For multibody final states the aim is clearly the understanding of \CP asymmetries in the interfering resonances rather than global asymmetries.
Of highest interest are those resonances that are closely related to the two-body modes used in $\Delta a_{\CP}$, for example the vector-pseudoscalar resonances $\Kstar\PK$ and $\Prho\Ppi$.
The measurement of further suppressed resonances is of interest as well since those have no contributions from gluonic penguin diagrams, thus allowing to constrain the source of \CP violating effects.

\subsection{\CP violation through triple product asymmetries}
In addition to rate asymmetries, whether global or local to a region of phase space, decays with a phase space, which is described by more than two degrees of freedom, can be used to study asymmetries in triple products.
Examples for triple products are (see \eg\ Ref.~\cite{Bigi:2009jj})
\begin{equation}
\vec{p_1}\cdot(\vec{p_2}\times\vec{p_3}),\qquad\vec{s}\cdot(\vec{p_1}\times\vec{p_2}),
\end{equation}
where $\vec{p_i}$ are the momenta of the decay products of \eg\ a four-body decay and $\vec{s}$ is a spin vector.

Such triple products are odd under parity transformation as well as under time reversal.
In the literature, these triple products are often referred to as $T$-odd variables, which indeed they are.
Others prefer just to refer to them as $P$-odd variables, since only decays are analysed and not their time reversal.
A detailed dissection of triple-products and derivative observables can be found in Ref.~\cite{Bevan:2014nva}.

In order to search for \CP violation, the asymmetry for the triple product $C_T$
\begin{equation}
a_T\equiv\frac{\Gamma(C_T>0)-\Gamma(C_T<0)}{\Gamma(C_T>0)+\Gamma(C_T<0)},
\end{equation}
is constructed from the partial rates of \PD decays with $C_T$ greater than or less than zero~\cite{Gronau:2011cf}.
Correspondingly, the asymmetry for the \CP conjugate decay is constructed as
\begin{equation}
\overline{a}_T\equiv\frac{\overline{\Gamma}(-\overline{C}_T>0)-\overline{\Gamma}(-\overline{C}_T<0)}{\overline{\Gamma}(-\overline{C}_T>0)+\overline{\Gamma}(-\overline{C}_T<0)},
\end{equation}
where $\overline{C}_T$ is the triple product for the corresponding \CP conjugate particles.

A non-zero value of the triple product asymmetries $a_T$ and $\overline{a}_T$ themselves is expected due to final state interactions (FSI)~\cite{Gronau:2011cf,Bigi:2001sg}.
However, FSI will lead to the same effect for $a_T$ and $\overline{a}_T$ and hence their difference is \CP asymmetric.
This leads to the definition of the \CP violating observable
\begin{equation}
a_{\CP}^{T-odd}\equiv\frac{1}{2}(a_T - \overline{a}_T),
\end{equation}
where the factor $1/2$ simply maintains the definition of an asymmetry.

Several experimental measurements of triple product asymmetries exist, notably including \decay{\Dz}{\Km\Kp\pim\pip} and \decay{\PD_{(s)}^+}{\KS\Kp\pim\pip} decays.
The most precise results exist for the former decay mode, for which the \babar collaboration measured $a_{\CP}^{T-odd}=(1.0\pm5.1\pm4.4)\times10^{-3}$~\cite{delAmoSanchez:2010xj} and the \lhcb collaboration measured $a_{\CP}^{T-odd}=(1.8\pm2.9\pm0.4)\times10^{-3}$~\cite{Aaij:2014qwa}.
In particular the latest \lhcb result shows the impressive robustness of triple product asymmetries against systematic uncertainties.
The current systematic uncertainty is dominated by the statistical uncertainty of a control sample with all other uncertainties being at the level of or less than $10^{-4}$.
This robustness is explained in the fact that flavour misidentification cannot generate a non-zero asymmetry by itself.
Furthermore, the resolution on $C_T$ only has to avoid a dilution around $C_T\approx0$ and particle identification has to avoid confusing the final state products.
Since these are not the most challenging requirements compared to other measurements, triple product asymmetries can be expected to remain limited by statistical precision in the foreseeable future and therefore an interesting way of searching for \CP violation.

\section{Rare charm decays}
\noindent Rare decays provide a wide range of interesting measurements.
The list of decay modes includes flavour-changing neutral currents, radiative, lepton-flavour violation, lepton-number violation, as well as baryon-number violation.
While a full discussion of rare charm decays would be beyond the scope of this paper a few remarks shall be made here.

\subsection{Rare decays and mixing}
\begin{figure}
\centering
\begin{tikzpicture}[scale=1.3,thick]
\node (charm) at (0,2) [] {\cquark};
\node (antiup) at (0,0) [] {$\bar{\uquark}$};
\node (decay) at (1,1) [vertex] {};
\node (invdecay) at (3,1) [vertex] {};
\node (up) at (4,2) [] {\uquark};
\node (anticharm) at (4,0) [] {$\bar{\cquark}$};
\node (Dzero) at (-0.3,1) [] {\Dz};
\node (Dzerobar) at (4.3,1) [] {\Dzb};
\path[-,decoration={markings,mark=at position .55 with {\arrow[draw=violet]{>}}}]
  (charm) edge[draw=violet,postaction={decorate}] (decay)
  (decay) edge [draw=violet,bend left=75,postaction={decorate}] node [above=3pt] {\mun} (invdecay)
  (invdecay) edge[draw=violet,postaction={decorate}] (up)
  (anticharm) edge[draw=violet,postaction={decorate}] (invdecay)
  (invdecay) edge [draw=violet,bend left=75,postaction={decorate}] node [below=3pt] {\mup} (decay)
  (decay) edge[draw=violet,postaction={decorate}] (antiup);
\end{tikzpicture}
\caption{Re-scattering diagram relating \decay{\Dz}{l^-l^+} decays to \Dz-\Dzb mixing.}
\label{fig:rare_rescatter}
\end{figure}
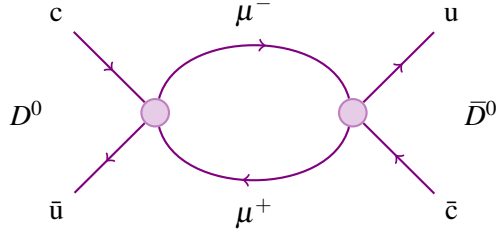

\noindent There is a direct link between mixing and flavour changing neutral current decays in several extensions of the standard model~\cite{Golowich:2007ka,Golowich:2009ii}.
These relate $\Delta C=1$ annihilation amplitudes, which mediate \decay{\Dz}{l^-l^+} decays, to $\Delta C=2$ mixing amplitudes where the annihilation product, \ie\ $l^-l^+$, creates a $\cquark\ubarquark$ pair from an initial \Dz meson (see Fig.~\ref{fig:rare_rescatter}).
At tree level, one example is a heavy $Z$-like boson with non-zero flavour-changing couplings.

The current central values of the \Dz-mixing parameters, notably $x$, translate into model-dependent limits for rare decays based on common amplitudes.
These rare decay limits lie significantly below the current experimental limits.
\Dz mixing is well established any upper limit from mixing will not change significantly in the future; however, the relevant parameter, $x$, remains to be measured with high precision.
Nevertheless, due to the direct correlation of mixing and rare decays, any observation above the model-dependent rare decay limits will rule out the corresponding model.
The best limit on flavour-changing neutral current decays is the recent \lhcb limit on the decay \decay{\Dz}{\mun\mup} of $6.2\times{}10^{-9}$ at $90\%$ confidence level\footnote{All further limits are given at $90\%$ confidence level as well.}~\cite{Aaij:2013cza}.

\subsection{Rare semi-leptonic decays}
\noindent 
Semi-leptonic decays involving exclusively charged or exclusively neutral leptons proceed via flavour-changing neutral current interactions, which are heavily suppressed in the standard model.
While the decays involving a neutrino anti-neutrino pair are beyond current experimental reach, those with a pair of charged leptons can be tested to very high precision.
The highest experimental precision is achieved in decays involving a $\mun\mup$ pair by the \lhcb collaboration, namely limits of $5.5\times10^{-7}$ for \decay{\Dz}{\pim\pip\mun\mup} decays~\cite{Aaij:2013uoa} and of $7.3\times10^{-8}$ ($4.1\times10^{-7}$) for \decay{\Dp(\Dsp)}{\pip\mun\mup} decays~\cite{Aaij:2013sua}.

\begin{figure}
\centering
\includegraphics[width=0.90\textwidth]{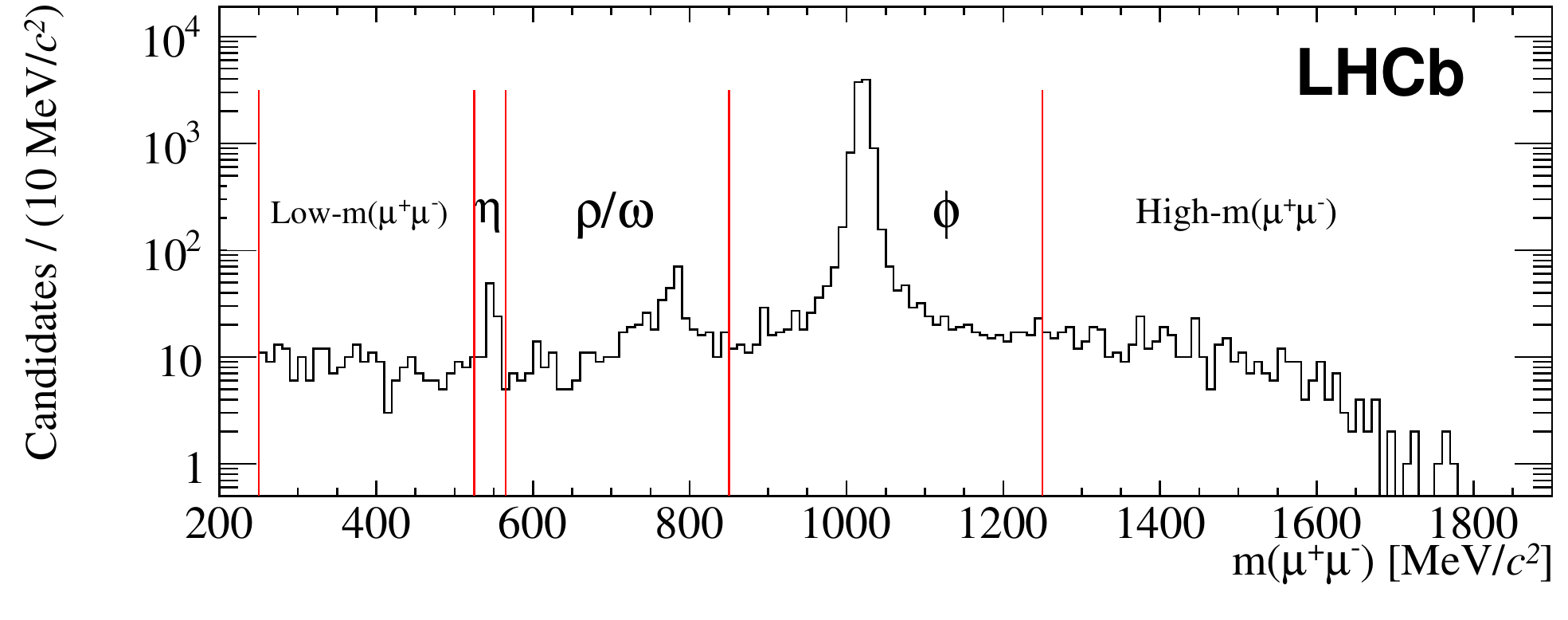}
\caption{$m(\mup\mun)$ spectrum of \decay{\PD^+_{(s)}}{\pip\mun\mup} candidates that pass the final selection and are within $3\sigma$ of either of the mean values of the Gaussian-like peaks describing the \Dp and \Dsp signals. Reproduced from Ref.~\cite{Aaij:2013sua}.}
\label{fig:rare_spectrum}
\end{figure}

At this level of precision it is necessary to exclude resonant regions that can contribute via allowed decays.
Figure~\ref{fig:rare_spectrum} shows this at the example of the $m(\mup\mun)$ spectrum of \decay{\PD^+_{(s)}}{\pip\mun\mup} candidates.
This distribution contains a contribution from \decay{\PD^+_{(s)}}{X\pip} decays, with $X=(\eta,\rho,\omega,\phi)$, followed by an allowed \decay{X}{\mun\mup} decay.
The remaining non-resonant distribution in Fig.~\ref{fig:rare_spectrum} is entirely due to background processes, but it is in these low-$m(\mup\mun)$ and high-$m(\mup\mun)$ regions that the search for non-resonant \decay{\PD^+_{(s)}}{\pip\mun\mup} has to be carried out.
The reported limit on the branching fraction is obtained from a limit in these two regions, which has been translated assuming a simple phase-space distribution for the non-resonant decay.

\subsection{Lepton-flavour, and lepton and baryon number violating decays}
\noindent Among lepton-flavour violating decays the most stringent constraint is a \belle search for \decay{\Dz}{\mump\epm} achieving a limit of $2.6\times{}10^{-7}$~\cite{Petric:2010yt}.
Searches for lepton-flavour violating muon or kaon decays already provide more constraining limits; however, in scenarious of non-universal couplings charm decays, giving access to the up-quark sector, are of great interest.

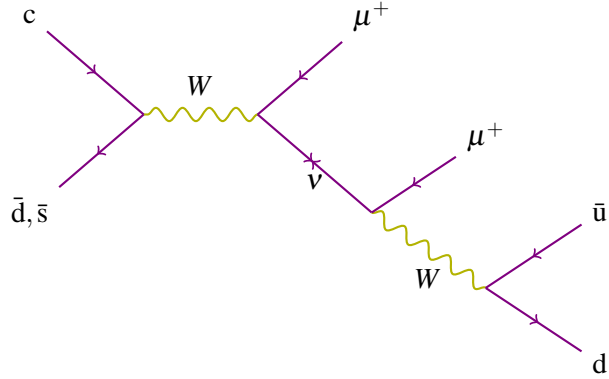
\begin{figure}
\centering
\begin{tikzpicture}[
        thick,
        level/.style={level distance=1.5cm},
        level 2/.style={sibling distance=2.6cm},
        level 3/.style={sibling distance=2cm}
    ]
    \coordinate
        child[grow=left]{
            child {
                node {\cquark}
                edge from parent [antielectron]
            }
            child {
                node {$\bar{\dquark},\bar{\squark}$}
                edge from parent [electron]
            }
            edge from parent [photon] node [above=3pt] {$W$}
        }
        child[grow=right, level distance=0pt] {
        child  {
            child {
                child {
                    node {$\dquark$}
                    edge from parent [electron]
                }
                child {
                    node {$\bar{\uquark}$}
                    edge from parent [antielectron]
                }
                edge from parent [photon] node [below=3pt] {$W$}
            }
            child {
                node {\mup}
                edge from parent [antielectron]
            }
            edge from parent [majorana]
            node [below] {$\nu$}
        }
        child {
            node {\mup}
            edge from parent [antielectron]
        }
    };
\end{tikzpicture}
\caption{
\label{fig:rare_majorana}
Feynman diagram of a \decay{\PD^+_{(s)}}{\pim\mup\mup} decay involving a Majorana neutrino.
}
\end{figure}

The best limit on lepton-number violating charm decays has been placed by the \lhcb collaboration on the decay \decay{\Dp}{\pim\mup\mup} with a limit of $2.2\times{}10^{-8}$~\cite{Aaij:2013sua}.
A particular interest in this type of decay lies in its implications on Majorana neutrinos.
If the neutrino is its own anti-particle, the leptonic \decay{\Dp}{\mup\nu} decay can be combined with a virtual anti-neutrino producing a positive muon and a negative pion via a $W^-$ boson (see Fig.~\ref{fig:rare_majorana}).

Only the \cleo collaboration has carried out searches for baryon-number violating charm decays.
Their best limit on the decay \decay{\Dz}{p\en} is $10^{-5}$ at $90\%$ confidence level~\cite{Rubin:2009aa}.
For a more complete overview of rare charm decays please refer to Ref.~\cite{Amhis:2014hma}.

\section{Conclusion}
\noindent Charm physics provides a vast range of insight into fundamental physics.
Clear signals for four-quark states including charm quarks have emerged.
The width difference of neutral charm mesons is firmly established.

Nevertheless, a number of open questions remain.
It has still not been firmly established whether charm mesons can oscillate into their anti-particles or not.
Flavour-changing neutral current decays remain elusive.
And of course the wholy grail of \CP violation, well known in the kaon and beauty sectors, remains to be found in charm particles despite experimental precision having reached levels better than $10^{-3}$.

With the \PB factories, \cleo-c and \cdf nearing completion of their data analysis, most new results will come from \lhcb and \besiii.
These are expected to explore very interesting territory for charm \CP violation.
The longer term future will be shaped by the \lhcb upgrade as well as future $\ep\en$ collider experiments, notably Belle II running at the beauty threshold, and possible future charm threshold machines.
Charm's third time has begun to produce its fruits, which may well develop into a real charm.

\section*{Acknowledgements}
\noindent The author would like to thank the organisers for the kind invitation to this interesting workshop, as well as M. Blanke, A. Buras, P. Krizan, and G. Tetlalmatzi-Xolocotzi for enlightening discussions.
Further thanks go to J. Brodzicka, E. Gersabeck, and S. Reichert for invaluable discussions on the manuscript of this paper as well as to J. Enegelfried for comments on an earlier version of this paper.

\clearpage

\bibliographystyle{utphys}       
\bibliography{gersabeck_charm_lecture}   

\end{document}